\documentclass[pra,amssymb,amsmath,twocolumn,floatfix]{revtex4}
\usepackage{psfrag}
\usepackage{mathrsfs}
\usepackage[colorlinks=true,linkcolor=blue,citecolor=blue]{hyperref}
\usepackage{enumerate}
\usepackage[dvips]{graphicx}
\usepackage{color}

\newcommand{\ra}{\rangle}
\newcommand{\la}{\langle}
\newcommand{\be}{\begin{equation}}
\newcommand{\ee}{\end{equation}}
\newcommand{\bea}{\begin{eqnarray}}
\newcommand{\eea}{\end{eqnarray}}
\newcommand{\ba}{\begin{array}}
\newcommand{\ea}{\end{array}}
\newcommand{\ben}{\begin{enumerate}}
\newcommand{\een}{\end{enumerate}}
\newcommand{\bi}{\begin{itemize}}
\newcommand{\ei}{\end{itemize}}
\newcommand{\bt}{\begin{table}}
\newcommand{\et}{\end{table}}
\newcommand{\btab}{\begin{tabular}}
\newcommand{\etab}{\end{tabular}}
\newcommand{\bfi}{\begin{figure}}
\newcommand{\efi}{\end{figure}}
\newcommand{\bfid}{\begin{figure*}}
\newcommand{\efid}{\end{figure*}}
\newcommand{\im}{\item}
\newcommand{\bd}{\begin{description}}
\newcommand{\ed}{\end{description}}

\newcommand{\diag}{\mathrm{diag}}
\newcommand{\nn}{\nonumber}

\newcommand{\mc}{\mathcal}
\newcommand{\mf}{\mathbf}
\newcommand{\bs}{\boldsymbol}

\newcommand{\ts}{\textsf}
\newcommand{\trm}{\textrm}
\newcommand{\tr}{\textrm{tr}}

\newcommand{\lb}{\label}

\newtheorem{res}{Result}

\usepackage{bm}
\newcommand{\finpr}{\hfill $\square$ \vspace{2mm}}

\def\be{\begin{eqnarray}}
\def\ee{\end{eqnarray}}
\def\bee{\begin{eqnarray*}}
\def\eee{\end{eqnarray*}}
\newtheorem{thm}{Theorem}

\newtheorem{lem}{Lemma}

          \def\tr{\hbox{Tr}}

\begin{document}

\title{Quantum algorithms for classical lattice models}

\author{G. De las Cuevas and W. D\"ur}
\affiliation{
Institut f{\"u}r Theoretische Physik, Universit{\"a}t Innsbruck, Technikerstra{\ss}e 25, A-6020 Innsbruck, Austria\\
Institut f\"ur Quantenoptik und Quanteninformation der \"Osterreichischen Akademie der Wissenschaften, Innsbruck, Austria}
\author{M. Van den Nest}
\affiliation{Max-Planck-Institut f\"ur Quantenoptik, Hans-Kopfermann-Str.~1, D-85748 Garching, Germany}
\author{M. A. Martin--Delgado}
\affiliation{Departamento de F\'{\i}sica Te\'orica I, Universidad Complutense, 28040 Madrid, Spain}

\begin{abstract}
We give efficient quantum algorithms to estimate the partition function of 
(i) the six vertex model on a two--dimensional (2D) square lattice, 
(ii) the Ising model with magnetic fields on a planar graph, 
(iii) the Potts model on a quasi 2D square lattice, and 
(iv) the $\mathbb{Z}_2$ lattice gauge theory on a three--dimensional square lattice. 
Moreover, we prove that these problems are \ts{BQP}--complete, that is, that estimating these partition functions is as hard as simulating arbitrary quantum computation. 
The results are proven for a complex parameter regime of the models. 
The proofs are based on a mapping relating partition functions to quantum circuits introduced in [Van den Nest \emph{et al.} Phys. Rev. A \textbf{80}, 052334 (2009)]  and extended here.
\end{abstract}
\date{\today} 

\maketitle


pacs: 03.67.-a, 03.67.Lx, 75.10.Hk, 75.10.Pq, 02.70.-c


\section{Introduction}
\label{sec:introduction}

Ising models are paradigmatic in analytical and numerical studies of phase transitions \cite{St71,Di95}. 
Their virtue is being simple enough to handle but nonetheless complex enough to capture the relevant physics. 
The same is true for other emblematic classical spin models, like the Potts model \cite{Wu84}, or the six vertex model \cite{Ba82}, which serve as toy models for certain physical systems. 
As a matter of fact, their applicability extends beyond physics, since spin models are used in the study of neural networks \cite{Me87}, biology \cite{Me87} and, more generally, statistical mechanical tools are also applied in economics \cite{Sc03}. 
The reason lies in the fact that these models study classical degrees of freedom (``spins'') which interact with each other (possibly in many--body interactions),
and this general scheme can serve as an abstract model for a wide class of systems. 

The central problem in the study of these models in equilibrium is the computation of their partition function $\mc{Z}:=\sum_{\mathbf{s}}e^{-\beta H(\mathbf{s})}$, where $H(\mathbf{s})$ is the Hamiltonian (or: ``energy function''), $\beta$ is defined as $\beta:=1/(k_BT)$ where $k_B$ is Boltzmann's constant and $T$ is the temperature, and $\mathbf{s}$ is the spin configuration. As is well known in statistical mechanics, this function captures all  relevant physical properties, since any thermodynamical quantity (such as the magnetization or the mean energy) can be derived as a function of $\mc{Z}$ \cite{Pa}.
In other words, knowledge of $\mc{Z}$ as a function of the parameters of the system amounts to complete knowledge of the thermal properties of the system. 
Thus, it is natural to investigate the computational complexity (colloquially speaking: the effort in terms of resources) required to compute or approximate $\mc{Z}$.
The emblematic models mentioned above have received most of the attention in this direction. 
For example, for the Ising model, Barahona \cite{Ba82} showed that computing its partition function in three dimensions (3D) is computationally very hard---to be precise, it is $\#$\ts{P}--complete, which is, colloquially speaking, the counting version of \ts{NP} \cite{Pa95}. 
This was an important contribution that settled the issue for those trying to tackle the problem after Onsager's success in solving the Ising model in two dimensions (2D) \cite{On44}. 

Further, one can raise the question of how hard it is to compute the partition function of these models on a \emph{quantum} computer. 
Quantum computers are known to offer a speedup over their classical counterparts in certain algorithms. Notably, the factoring problem is known to be feasible by a quantum computer \cite{Sh94}, while it remains intractable in all known classical algorithms. 
Roughly speaking, `feasible' means that the resources (in time and space) required to solve it scale polynomially with the size of the input of the problem, and intractable means that they may scale exponentially. 
The quantum computational complexity of classical spin models has been addressed, for example, in \cite{Li97}, where a quantum algorithm for the Ising partition function was presented (see also \cite{Li04,Ge08,Ger08,Ge10,Br07} for related work).
Further, in \cite{Ah07} it was proven that computing the partition function of the Potts model is \ts{BQP}--complete (see also \cite{Ah06,Wo08}). 
\ts{BQP} stands for bounded--error quantum polynomial time, and, colloquially speaking, is the class of decision problems that can be efficiently approximated by a quantum computer, and the hardest problems in this class are called \ts{BQP}--complete. 
This situation contrasts with that of a different class of classical spin models, namely lattice gauge theories with gauge group $\mathbb{Z}_2$ \cite{Ko79}, for which, to the best of our knowledge, no results are known concerning their quantum computational complexity. These are models with ``Ising variables'' (i.e. classical degrees of freedom with two states) but which nonetheless exhibit local symmetries~\cite{We71,Wi75}. 

In this paper we tackle the question of the quantum computational complexity both of several paradigmatic classical spin models as well as of a $\mathbb{Z}_2$ lattice gauge theory. Our approach builds upon a mapping between partition functions and quantum circuits introduced in \cite{Va09}. In that work, this mapping is exploited to show, among others, that estimating the partition function of the six vertex model and of Ising--type models is \ts{BQP}--complete. 
Here we revisit this approach and extend the mapping to standard Ising models, Potts models and to $\mathbb{Z}_2$ lattice gauge theories.
Based on that, we provide efficient quantum algorithms to estimate (with polynomial accuracy) the partition function of 
the six vertex model on a 2D square lattice, 
the Ising model with magnetic fields on a planar graph, 
the Potts model on a quasi 2D square lattice, 
and a 3D lattice gauge theory with gauge group $\mathbb{Z}_2$.
Moreover, we show that computing the partition functions of these models is \ts{BQP}--complete, that is, it is as hard as simulating arbitrary quantum computation.  
Therefore, in this work we put paradigmatic classical spin models and $\mathbb{Z}_2$ lattice gauge theories on an equal footing as far as their quantum computational complexity is concerned. 

However, a word of caution is needed here: our results are valid mostly for a \emph{complex} parameter regime of the models. This means that we can prove the complexity results only if the coupling strengths of these models have certain imaginary values (this problem is also encountered in \cite{Va09,Ah07}). 
Note that although such complex parameters do not correspond to physical models, the partition function with complex arguments is commonly studied, e.g., in the context of evaluating the Tutte polynomial or finding (complex) zeros of $\mc{Z}$ to identify phase transition points \cite{So05}.
We also want to stress that our results rely crucially on the fact that an \emph{additive} approximation of the partition function is obtained (in contrast with a multiplicative approximation or an exact calculation), which, roughly speaking,  means that the desired quantity is approximated with polynomial accuracy. 

To summarize, the main results of this work are the following. 

\textbf{Statement of results}. 
\emph{
Consider the partition function $\mc{Z}$ of the following classical spin models
\ben[(i)]
\im the six vertex model defined on a rectangular grid
\im the Ising model with magnetic fields defined on a planar graph,
\im the three--level Potts model on a quasi 2D square lattice  with certain boundary conditions, and 
\im the $\mathbb{Z}_2$ lattice gauge theory on a 3D square lattice,
\een 
defined on a certain complex parameter regime, that is, the value of coupling strength $J$ is, for example, $e^{\beta J} = i$. 
Furthermore, consider that there are certain values of the coupling strengths which appear together (i.e.~certain sets of neighboring spins whose interaction takes a specific value). 
Then we provide efficient quantum algorithms to approximate the partition function $\mc{Z}$ of these models with polynomial accuracy. 
Moreover, we show that estimating these partition functions is \ts{BQP}--complete, that is, it is as hard as simulating arbitrary quantum computation. 
}

Finally, as an extension of our results, we show that, if the Ising model considered above is defined on a lattice with periodic boundary conditions, then estimating its partition function with polynomial accuracy is \ts{DQC1}--hard.
This is the representative class of a scheme for quantum computation called the `one clean qubit model', where all qubits but the first one are initialized in a totally mixed state~\cite{Kn98}. 

This paper is structured as follows. 
In Sec.~\ref{sec:classical} we give a background on classical spin models, as well as on some notions of complexity of classical spin models. 
Then we review how to define \ts{BQP}--complete problems as an estimation of a unitary matrix element (Sec.~\ref{sec:matrix_elements_BQP}). 
In Sec.~\ref{sec:mappings} we show how to relate partition functions of classical spin models with spin circuits, based on \cite{Va09} and extended here. 
The main results of this work are presented in Sec.~\ref{sec:BQP}, where we show the \ts{BQP}--completeness of the six vertex model, the Ising model, the Potts model, and of the 3D lattice gauge theory with gauge group $\mathbb{Z}_2$ with certain conditions. 
In Sec.~\ref{sec:further} we present an extension of the results related to the one--clean--qubit model. 
Finally, we present our conclusions in Sec.~\ref{sec:conclusions}.

\section{Classical spin models}
\lb{sec:classical}

In this section we will present some general considerations about classical spin models as well as some facts concerning their computational complexity. 

\subsection{General background}
\lb{ssec:background}

Classical spin models have proven successful to model magnetism, where they capture interesting physics such as critical phenomena despite their simplicity. More generally, classical spin models can serve as toy models for complex systems. For example, Ising models have been used to model neural networks (see Hopfield networks), or spin glasses (see, e.g., \cite{Me87}). 

Let us now define what is understood by a classical spin model. Common to all of them are the following ingredients:
\ben[(i)]
\im
A  degree of freedom represented by a classical spin which may take on a set of values: $s \in \{0,1,...,q-1\}$. This is a $q$--level system.
\im 
A lattice, or more generally an arbitrary graph $G$, to which the classical spins are associated. They can sit at the vertices, edges or faces of the graph; the idea is that the graph encodes the interaction pattern of the model.  The complexity of the model is affected by the lattice or graph on which it is defined, as we will see below.
\im
An energy function $H(\mathbf{s})$ depending on a given spin configuration and the coupling strengths representing the types of interactions of the model: nearest--neighbor, many--body interactions, magnetic fields, etc.
\im
A partition function $\mc{Z}$ which is obtained by summing over the Boltzmann weights of all spin configurations, $\mc{Z}=\sum_{\mathbf{s}}e^{-\beta H(\mathbf{s})}.$
\een

From this common structure, several different families of models can be distinguished. 
One of the most relevant criteria is whether the model exhibits global or local symmetries (if any). We shall refer to the former as standard statistical models and to the latter as lattice gauge theories. This distinction of symmetry has profound consequences in the physics of the models. It is also related to their range of applicability: standard statistical models appear naturally in descriptions of classes of condensed matter systems and lattice gauge theories originated in the study of the fundamental interactions in nature and elementary particles.

Dimensionality is another key distinction which determines the complexity of the models. However, note that in order to have a well--defined notion of dimension for a graph $G$, this must be embedded in a smooth manifold of dimension $D$.

Standard statistical models can be divided into two big families depending on where the interactions on the graph $G$ take place: vertex models and edge models. Vertex models were introduced to describe ice--type models, crystals with hydrogen bonding or ferroelectrics \cite{Ba82}. Edge models were introduced to explain phase transitions in materials with elementary magnetic moments \cite{Mai03,Ni11}. 

A vertex model consists of classical spins, namely $q$--level particles $s_e \in \{0,1,\ldots ,q-1\}$, which are placed on the \emph{edges} of the lattice, and (typically many--body) interactions take place on the vertices. In the case of a (tilted) 2D lattice, one deals with four--body interactions between neighboring particles, and each of the spins participates in only two interactions. The Hamiltonian of such a system is given by
\be
H= \sum_{a \in V} h^a (s_i,s_j,s_k,s_l) \, ,
\ee
where $h^a (s_i,s_j,s_k,s_l)$ is a (local) four--body interaction term between the spins $s_i,s_j,s_k,s_l$ sitting at the edges incident to vertex $a$. 
We denote the Boltzmann weight associated with this local energy as 
\be
w^{a}(s_i,s_j,s_k,s_l) : = e^{-\beta h^{a}({s_i,s_j,s_k,s_l})}\, .
\lb{eq:wa}
\ee
The partition function is obtained by multiplying all local Boltzmann weights, and summing these over all spin configurations $\mf{s}$,
\be
\mc{Z}_{\mbox{\scriptsize{vm}}} = \sum_{\mf{s}} \prod_a w^a (s_i,s_j,s_k,s_l)\, .
\ee

In contrast to vertex models, in edge models the classical spins sit at the vertices of the graph, also taking $q$ possible states, $s_i\in\{0,\dots, q-1\}$, and interactions take place along the edges. 
Consider, thus, a $q$--state edge model on an $n\times m$ square lattice with an edge--dependent energy function $h^e(s_i, s_j)$. 
Let 
\be
w^e(s_i, s_j):=e^{-\beta h^e(s_i, s_j)}
\ee
denote the corresponding Boltzmann weight. Then the partition function is given by ${\cal Z}= \sum_{\mf{s}} \prod_{e=ij} w^e(s_i, s_j)$. 

Concerning Lattice Gauge Theories (LGTs), the most relevant criterion to classify them is whether the 
internal gauge group ${\cal G}$ is Abelian (discrete like ${\cal G}=\mathbb{Z}_q$,
or continuous ${\cal G}=U(1)$), or non--Abelian (discrete such as
a permutation group ${\cal G}=S_3$, or continuous such as ${\cal G}=SU(N)$).
We refer the reader to \cite{Ko79} for an introduction to these models.
In this work we will focus on $\mathbb{Z}_2$ LGTs, and we will be in the following features: they are models whose classical spins can take two values $s_e\in\{0,1\}$, they sit at the edges of a $d$--dimensional square lattice, and they interact along the faces of this lattice. 
More precisely, the interaction of spins $s_i,s_j,s_k,s_l$ at the boundary of face $f$, $\partial f$, has the form
\be
h_f(s_i, s_j, s_k,s_l) = -J_f\delta(s_i+ s_j + s_k +s_l)\, ,
\lb{eq:int-LGT}
\ee
where the sums are performed modulo 2 throughout this section, and $\delta(0)=1$ and it is 0 otherwise.
The Hamiltonian is then obtained as a sum over interactions on every face:
\be
H(\mf{s}) = -\sum_f h_f (\{s_e:e \in \partial f\})\, ,
\ee
where $\partial f$ denotes the boundary of face $f$.

These models exhibit $\mathbb{Z}_2$ gauge symmetry; more precisely, its Hamiltonian is invariant under $\mathbb{Z}_2$ operations around any vertex, $g_v = \prod_{e\in \trm{inc } v}X_e$, where $e\in \trm{inc } v$ denotes all edges incident to vertex $v$, and $X_e$ is a flip operator, $X_e: s\to s+1$.
One can use this symmetry to eliminate some degrees of freedom, a process usually referred to as `gauge fixing'. A specific choice of this fixing is the `temporal gauge', where all degrees of freedom in one particular direction (the one associated with time) are fixed. A restriction about gauge fixing that concerns us is the fact the edges whose variable has been fixed by the gauge cannot form a closed loop~\cite{Cr77}. This fact will be important in our proof of the \ts{BQP}--completeness of this model in Sec.~\ref{ssec:3DZ2LGT}. 
 
\subsection{Classical computational complexity of spin systems}
\lb{ssec:computational-complexity}

In this section we will be interested in the classical computational complexity of the classical spin models presented in Sec.~\ref{ssec:background}. Generally speaking, understanding the properties of, say, the Ising model on some graph is a difficult task. This is reflected by the fact that the Ising and other models are associated with hard problems in computational complexity theory \cite{Ba82}. For concreteness we will focus in the following on the Ising model, but the considerations in this section have general relevance.

Most prominently, the Ising model is known to be associated with computational problems which are \ts{NP}--complete; here \ts{NP} stands for `non-deterministic polynomial time'. The complexity class \ts{NP} consists of all decision problems $f$ (i.e.~YES/NO questions) which have the property that, for every input $x$ for which it is claimed that $f(x) = $ YES, there exists a `short proof' that this is indeed the case, i.e.~a proof  which may be efficiently verified as a function of the size of the input. More precisely, for each problem $f\in $ \ts{NP} it is required that there exists an efficiently computable function $V$ (the verifier of the proof) such that:
\begin{itemize}
\item[] For every input $x$, one has $f(x) = $ `YES' if and only if there exists a poly--size bit string $\xi$ (the witness) satisfying $V(x, \xi) = $ `YES'.
\end{itemize}
Colloquially speaking,  \ts{NP} problems have the property that, whenever a solution to the problem is \emph{proposed} (e.g.~by an untrusted third party), it is possible to efficiently \emph{verify} whether this proposed solution is indeed correct. Note that in the definition of \ts{NP} no mentioning is made of the difficulty of \emph{finding} a solution; even though a problem has an efficient verifier, it is a priori not excluded that the time required to find a solution scales exponentially with the input size.

An archetypical \ts{NP} problem related to the Ising model is the problem of deciding whether the ground state energy of $H_G(\mathbf{s})$ on a graph $G$ (which constitutes the input of the problem) is below a certain value $K$. This problem is indeed in \ts{NP}: if the ground state energy of $H(\mathbf{s})$ is smaller than $K$, then the ground state provides a witness which allows to efficiently verify this fact; the verifier function is nothing but the energy function  $H_G(\mathbf{s})$.

Not only is the problem of determining the Ising ground state in \ts{NP}, it is among the hardest problems in this complexity class. This is reflected by the fact that this problem is known to be \ts{NP}--complete. This means that every problem in the class \ts{NP} can be reduced, with only polynomial computational effort in the input size of the problem, to an instance of the Ising ground state problem. This implies, in particular, that the existence of an efficient algorithm for the Ising ground state problem would yield an efficient algorithm for all problems in \ts{NP}. The \ts{NP}--completeness of the Ising model thus points to an intrinsic computational difficulty of this simple system.

There are several variants of the Ising ground state problem which are known to be \ts{NP}--complete. We mention two of them.

\begin{thm}\cite{Ba82} \label{thm_Ising_NP}
The following problems are \ts{NP}--complete:
\ben[(i)]
\im Given a graph $G$ and an integer $K$, determine whether the ground state energy of $H(\mathbf{s})=\sum_{e=ab} s_as_b$ is smaller than $K$.

\im Given a planar 
\footnote{A planar graph is a graph which can be drawn in the plane without crossings of the edges.} 
graph $G$ and an integer $K$, determine whether the ground state energy of $H(\mathbf{s})= \sum_{e=ab}  s_as_b + \sum_{a} s_a$ is smaller than $K$.
\een

\end{thm}

We remark that, in the second of these problems, the presence of the external fields ($h_a\equiv -1$) is crucial to obtain \ts{NP}--completeness. Indeed, it is known that the ground state energy, as well as the partition function, of the Ising model on an arbitrary planar graph without external fields can be efficiently computed. We also note that the quantum computational complexity of the Ising model, as will be discussed in Sec.~\ref{ssec:Ising}, will involve Ising models on planar graphs in the presence of magnetic fields.

The \ts{NP}--completeness of the above ground state problems has strong implications for the evaluation of the corresponding partition functions. First, once the partition function of a model can be evaluated efficiently, also the ground state energy of the model can be efficiently determined: the evaluation of the partition function is `at least as hard' as the evaluation of the ground state energy. Consequently, for the \ts{NP}--complete Ising models, the evaluation of their partition function is  \ts{NP}--hard, i.e.~at least as hard as any problem in \ts{NP}. Note, however, that the evaluation of the partition function does not belong to the class \ts{NP}, as it is not a decision problem but rather a counting problem.  The relevant complexity class in this case is $\#$\ts{P} (`sharp-\ts{P}'). Given an efficiently computable decision problem $g$ (i.e.~$g\in $ \ts{P}), the problem of determining how many inputs yields the answer `YES', represented by the number $\#g = |\{ x : g(x) = `\mbox{YES}$'$\}|$, defines the complexity class $\#$\ts{P}.

Since the ground state problems of the Ising models in Theorem~\ref{thm_Ising_NP} are \ts{NP}--complete, it can be shown that computing the corresponding partition functions  are $\#$\ts{P}--complete problems: every problem in $\#$\ts{P} can be reduced, with polynomial computational effort, to the evaluation of the partition function of such an Ising model on some graph. This is formulated in the following result:

\begin{thm}\cite{Ba82}
The following problems are $\#$\ts{P}--complete:
\begin{itemize}
\item[(i)] Given a graph $G$ and $\lambda=e^{-\beta}$, determine the partition function ${\cal Z}(\lambda)$ of the Ising model on $G$ with energy $H(\mathbf{s})=\sum_{e=ab} s_as_b$.
\item[(ii)] Given a planar graph $G$ and $\lambda=e^{-\beta}$, determine the partition function ${\cal Z}(\lambda)$ of the Ising model on $G$ with $H(\mathbf{s})= \sum_{e=ab}  s_as_b + \sum_{a} s_a$.
\end{itemize}
\end{thm}

\section{Unitary matrix elements and \ts{BQP}--completeness}
\label{sec:matrix_elements_BQP}

The goal of this paper is to relate the evaluation of partition functions to problems which are complete for \emph{quantum} complexity classes. The main complexity class which will be considered is `bounded-error quantum polynomial time' (\ts{BQP}), representing the class of decision problems which can be solved efficiently on a quantum computer. The route we will take to prove \ts{BQP}--competeness of certain partition function problems will be to start from a standard complete problem for \ts{BQP} and then to relate these problems to the approximation of  partition functions. The standard \ts{BQP}--complete problem in question involves estimating matrix elements of unitary quantum circuits, which we briefly discuss here.

Consider a quantum circuit $U$ acting on $n$ qubits, which is composed of poly$(n)$ gates acting each on, say, at most two qubits. Let $\{|0\rangle, |1\rangle\}$ denote the single--qubit computational basis. Then there exists a well known technique to estimate the matrix element $\langle 0 |^{\otimes n}U|0\rangle^{\otimes n}$ in poly--time on a quantum computer, using the `Hadamard test'
(see Fig.~\ref{fig:Hadamardtest} and its caption for a brief a explanation---we refer to the literature, e.g.~\cite{Ah06,Ah06b,Wo08,Sm10}, for further explanations). More precisely, for any approximation scale $\epsilon$ which scales at most inverse polynomially with $n$, the Hadamard test returns a complex number $c$ which satisfies
\be 
|c - \langle 0 |^{\otimes n}U|0\rangle^{\otimes n}| \leq \epsilon,
\ee
with a success probability that is exponentially (in $n$) close to $1$. 

\bfi[htb]\centering
\psfrag{M}{$0,1$}
\psfrag{0}{$|0\ra$}
\psfrag{H}{$H$}
\psfrag{U}{$U$}
\includegraphics[width=0.75\columnwidth]{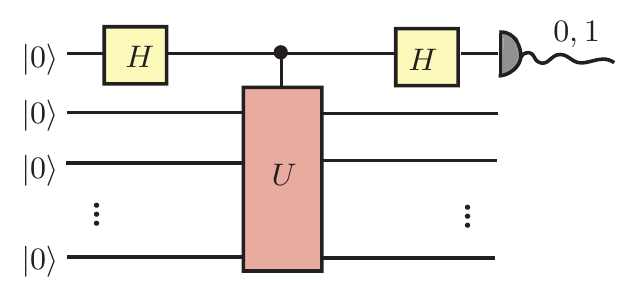}
\caption[Hadamard test.]{The Hadamard test. This is performed with a quantum circuit where the first qubit is transformed with a Hadamard gate, then the unitary $U$ is applied to rest of the qubits conditional on the state of the first qubit, and then a Hadamard gate is applied to the first qubit again. Finally, this qubit is measured in the $\sigma_z$ basis. 
From the probability to obtain 0 and 1, $p_0$ and $p_1$ respectively, one can estimate the real part of $c$ (denoted $\trm{Re}(c)$), as they are related by $p_0= [1+\trm{Re}(c)]/2$ and  $p_1= [1-\trm{Re}(c)]/2$. 
To estimate the complex part,  
the circuit is modified to include a phase gate $P= \trm{diag}(1,i)$ after the first Hadamard gate and before the controlled--$U$ gate. 
The probabilities to measure 0 and 1 are then related to the imaginary part of $c$ (denoted $\trm{Im}(c)$) as $\tilde{p_0}=[1-\trm{Im}(c)]/2$ and $\tilde{p_1}=[1+\trm{Im}(c)]/2$ 
(see also \cite{Ah06,Ah06b,Wo08,Sm10}).
}
\lb{fig:Hadamardtest}
\efi

Moreover, estimating the above matrix element problem is \ts{BQP}--hard, i.e.~every decision problem that can be solved efficiently with a quantum computer can be reduced, with polynomial (classical) computational effort, to the estimation of such a matrix element with the aforementioned accuracy $\epsilon$. 
Without loss of generality we can restrict all operations to act on nearest--neighboring qubits, since the SWAP operation can be used to move distant qubits to nearest--neighbor positions (with linear overhead in the number of qubits). Thus, one has the following. 
\begin{thm}\label{thm_matrix_elements}
The following problem is \ts{BQP}--hard:
\begin{itemize}
\item[] Consider an $n$--qubit quantum circuit $U$ consisting of a polynomial number of two--qubit gates acting on nearest--neighbor qubits.  Then provide a number $c$ such that 
\be 
|c - \langle 0 |^{\otimes n}U|0\rangle^{\otimes n}| \leq \frac{1}{\mbox{poly(n)}}
\ee 
holds with a probability which is exponentially (in $n$) close to $1$.
\end{itemize}
\end{thm}

We provide a few remarks:
\ben[(i)]
\im One may adapt the formulation of Theorem~\ref{thm_matrix_elements} in a straightforward way to arrive at a \ts{BQP}--complete decision problem (i.e.~a problem that is both \ts{BQP}--hard and in \ts{BQP}). To do so, one considers circuits $U$ for which it is \emph{promised} that $|\langle 0 |^{\otimes n}U|0\rangle^{\otimes n}|$ is either $\leq 1/3$ or $\geq 2/3$, and the goal is to decide which of these two cases holds. 

\im 
\ts{BQP}--hardness of the matrix element problem in Theorem~\ref{thm_matrix_elements} is maintained if one considers quantities of the form $\langle \psi|U|\psi'\rangle$, where $|\psi\rangle$ and $|\psi'\rangle$ are \emph{fixed} complete product states,
instead of $\langle 0 |^{\otimes n}U|0\rangle^{\otimes n}$. This is because instances of the former matrix element problem can easily be reduced to instances of the latter. This property will be used in Secs.~\ref{ssec:6VM}, \ref{ssec:Ising} and \ref{ssec:Potts}.

\im 
\ts{BQP}--hardness is maintained when the circuit $U$ is restricted to be composed of gates from a strongly universal elementary gate set. An elementary gate $S$ consisting of gates acting on at most two qubits is said to be strongly universal if every two--qubit unitary operation can be approximated with arbitrary accuracy by products of elements from $S$. Further, \ts{BQP}--hardness is also maintained if, instead of considering strongly universal unitary gate sets, gate sets are considered that are \emph{encoded} universal for quantum computation, for a suitable notion of encoded universality. This will be important for our proofs in Secs.~\ref{ssec:6VM}, \ref{ssec:Potts} and \ref{ssec:3DZ2LGT}.
\een

\section{Mappings between classical lattice models and quantum circuits}
\lb{sec:mappings}

In this section we review the mappings between classical lattice models and quantum circuits (introduced and used in \cite{Va09}), and we will extend them to standard Ising models, to Potts models and to $\mathbb{Z}_2$ LGTs.
The mappings will allow us to interpret the partition function of the classical model as a quantum expectation value. 
More precisely, consider a circuit ${\cal C}$ consisting of (unitary) quantum gates. 
Then we will show that partition functions ${\cal Z}$ of vertex models, edge models and $\mathbb{Z}_2$ LGTs are related to matrix elements of certain quantum circuits ${\cal C}$ as 
\be
{\cal Z} =  \kappa \: \la L| {\cal C} | R\ra \, ,
\lb{eq:Z=LCR}
\ee
where $\la L|$ ($|R\ra$) are product states determined by the left (right) boundary conditions of the classical spin model (see below), and $\kappa$ is a constant depending on the size of the lattice, i.e.~the number of classical spins.

\subsection{Vertex models}
\label{ssec:vertex-models}

We start by considering vertex models (see Sec.~\ref{ssec:background}).  Here we essentially follow the argument of \cite{Va09}. For illustration purposes, we will concentrate on a tilted 2D square lattice. However, our mappings are not restricted to such lattices and are easily generalized to other (regular) lattices.

Our construction begins with the following observation. The local Boltzmann weights \eqref{eq:wa} associated to each interaction can be seen as rank--four tensors of dimension $q$, or, equivalently, as $q^2 \times q^2$ matrices by grouping the indices into left indices $(i,j)$ and right indices $(k,l)$. We take the latter approach to associate each of these matrices with a quantum gate acting on two $q$--level quantum states (see Fig.~\ref{fig:VM}), 
\be
W^a:=\sum_{s_i,s_j,s_k,s_l} w^a(s_i,s_j,s_k,s_l) |s_i,s_j\ra \la s_k,s_l|\, .
\lb{eq:Wa}
\ee
Note that the right indices $(k,l)$ correspond to the input of the gate, while the left indices $(i,j)$ represent the output. The state is thus processed from right to left, where gates corresponding to outermost right vertices are performed in the first place 
\footnote{This is opposite to the way quantum circuits are usually drawn, where the first processing takes place from left to right.}. 
The corresponding quantum circuit ${\cal C}$ is given by $m$ layers of nearest--neighbor two--qubit gates (see Fig.~\ref{fig:VM}),
\be
\mc{C} = \prod_a W^a \, .
\ee
This can be seen as the contraction of a tensor network, i.e.~as a summation over joined indices. For translational invariant models, each of the layers corresponds to the transfer matrix \cite{Ba82} of the classical model.

Thus, generally speaking, we have mapped a product of interactions of the classical spin model (which is essentially a partition function) to a contraction of quantum gates (which is essentially a quantum circuit). 
Now we only need to show how to map the boundary spins. 
We consider an $n \times m$ lattice with fixed boundary conditions, namely a lattice whose spins at the left and right boundary are fixed in some arbitrary configuration $L = (s_1^L, \dots, s^L_n)$ and $R=(s^R_1, \dots, s^R_n)$, respectively. 
Using the spin states as computational basis states, we map these to two $n$--particle quantum states:
\be
\ba{rcl}
\la L | &=& \la s_1^L|  \ldots \la s^L_n| \, ,\\
|R\ra &=& |s_1^R\ra \ldots |s_n^R\ra\, .
\ea
\lb{eq:LR}
\ee
where we omit the tensor product symbol throughout this paper. 
Note that, since the qubits are processed from right to left, the state $|R\ra$ serves as input for the circuit, while $|L\ra$ constitutes the readout basis state. 

It is now straightforward to see that the overlap of the resulting state ${\cal C}|R\rangle$ with the product state $\langle L|$ is exactly the partition function of the classical vertex model,
\be
\label{partitionVM}
{\cal Z}_{\mbox{\scriptsize{vm}}}^{L,R} = \langle L|{\cal C}|R\rangle	\, .
\ee
This concludes the mapping between matrix elements of quantum circuits and partition functions of classical spin models. 

\bfi[htb]\centering
\psfrag{a}{\small{$s_1^L$}}
\psfrag{b}{\small{$s_2^L$}}
\psfrag{c}{\small{$s_n^L$}}
\psfrag{d}{\small{$s_1^R$}}
\psfrag{e}{\small{$s_2^R$}}
\psfrag{f}{\small{$s_n$}}
\psfrag{i}{\small{$i$}}
\psfrag{j}{\small{$j$}}
\psfrag{k}{\small{$k$}}
\psfrag{l}{\small{$l$}}
\psfrag{w}{\small{$w^a(s_i,s_j,s_k,s_l)$}}
\psfrag{g}{\small{$\la s_1^L|$}}
\psfrag{h}{\small{$\la s_2^L|$}}
\psfrag{o}{\small{$\la s_n^L|$}}
\psfrag{u}{\small{$W^a$}}
\psfrag{1}{\small{$i$}}
\psfrag{2}{\small{$j$}}
\psfrag{3}{\small{$k$}}
\psfrag{4}{\small{$l$}}
\psfrag{p}{\small{$| s_1^R\ra$}}
\psfrag{q}{\small{$| s_2^R\ra$}}
\psfrag{r}{\small{$| s_n^R\ra$}}
\psfrag{x}{\small{$s_1^L$}}
\psfrag{y}{\small{$s_2^L$}}
\psfrag{z}{\small{$s_3^L$}}
\psfrag{X}{\small{$s_1^R$}}
\psfrag{Y}{\small{$s_2^R$}}
\psfrag{Z}{\small{$s_3^R$}}
\psfrag{I}{\small{$i$}}
\psfrag{J}{\small{$j$}}
\psfrag{K}{\small{$k$}}
\psfrag{m}{\small{$w_{ij}$}}
\psfrag{n}{\small{$w_{jk}$}}
\psfrag{5}{\small{$i$}}
\psfrag{6}{\small{$j$}}
\psfrag{7}{\small{$k$}}
\psfrag{M}{\small{$w_{ij}$}}
\psfrag{N}{\small{$w_{jk}$}}
\psfrag{G}{\small{$\la s_1^L|$}}
\psfrag{H}{\small{$\la s_2^L|$}}
\psfrag{O}{\small{$\la s_n^L|$}}
\psfrag{P}{\small{$| s_1^R\ra$}}
\psfrag{Q}{\small{$| s_2^R\ra$}}
\psfrag{R}{\small{$| s_n^R\ra$}}
\psfrag{t}{\small{time}}
\psfrag{A}{(a)}
\psfrag{B}{(b)}
\hspace{-4mm}\includegraphics[width=1\columnwidth]{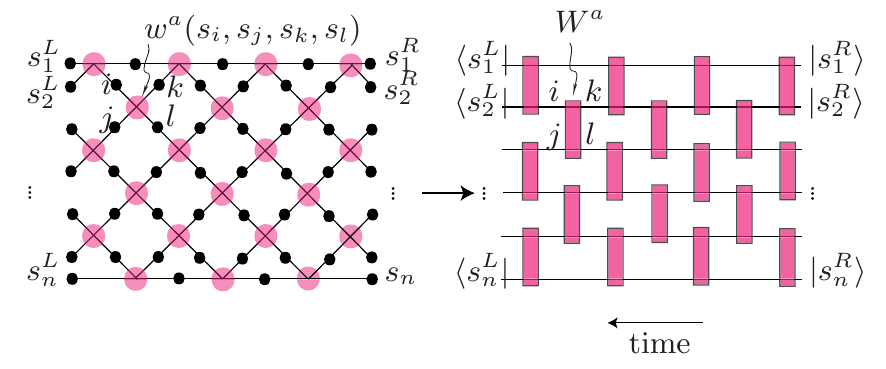}
\caption{
(Left) In a vertex model, particles (black dots) sit at the edges and interactions (pale red dots) take place in the vertices. (Right) This model is mapped to a quantum circuit, where each particle becomes a qubit, and each interaction a two--qubit gate.
}
\label{fig:VM}
\efi

Now we make some remarks concerning this mapping. 
\ben[(i)]
\im\lb{im:vertex-complex}
\emph{Complex couplings.} Note that the quantum gates are specified by the parameters of the classical model, namely by the Boltzmann weights of the local interactions. It follows that the gates $W^a$ (Eq.~\eqref{eq:Wa}) are unitary only in certain parameter regimes of the  interaction $w^a$ of the classical spin model.
This will lead to the requirement of complex parameters in our proof of Sec.~\ref{ssec:6VM}.
\im\lb{im:vertex-OPBC}
\emph{Open and periodic boundary conditions.}
This mapping can be easily extended to open boundary conditions, i.e.~to systems where the left and right spins are `free' and thus fully summed out in the partition function. 
This is achieved by replacing the left and right states $|L\rangle$ and $|R\rangle$ by the state $|+\rangle^{\otimes n}$, where $|+\rangle= q^{-1/2} \sum_{i=0}^q |i\rangle$ is a superposition over all $q$ single spin states. This gives rise to the identity 
\be 
\label{Z_free} 
{\cal Z}_{\mbox{\scriptsize{vm}}}^{\small{\trm{OBC}}} = q^n\langle +|^{\otimes n}{\cal C}|+\rangle^{\otimes n}\, .
\ee
Furthermore, periodic boundary conditions can also be taken into account by summing over the diagonal matrix elements, which results in
\be 
{\cal Z}_{\mbox{\scriptsize{vm}}}^{\trm{PBC}} =\tr({\cal C})\, .
\ee

\im\lb{im:vertex-geometries}
\emph{Other geometries.}
One can also consider other geometries such as vertex models on a 2D tilted triangular lattice \cite{Sa75}, where six--body interactions take place at the vertices and the Boltzmann weights can be arranged into $q^3 \times q^3$ matrices, corresponding to a quantum circuit with three--body quantum gates. One such model is the 32 vertex model \cite{Sa75}. Also 3D models, such as models on a tilted 3D square lattice, are of this type and can thus be mapped to quantum circuits. In this case, one deals with three--body gates acting on a 2D array of quantum particles. 
\een

\subsection{Edge models}
\label{ssec:edge-models} 

In the following we will present a mapping for edge models, also following \cite{Va09}. 
Unlike vertex models, in edges models interactions take place at the edges, as explained in Sec.~\ref{ssec:background}. For the mapping we will distinguish between interactions at horizontal or vertical edges. More precisely, we associate a $q\times q$ matrix to each horizontal edge $e$: 
\be
\label{eq:W_e_horizontal} 
W_e^h:= \sum_{s_i, s_j} w^e(s_i, s_j) |s_j\rangle\langle s_i|\, ,
\ee 
and a $q^2\times q^2$ \emph{diagonal} matrix to each vertical edge $e$, 
\be
\label{eq:W_e_vertical} 
W_e^v:= \sum_{s_i, s_j} w^e(s_i, s_j) |s_i, s_j\rangle\langle s_i, s_j|\, .
\ee 
The matrices $W_e^h$ and $W_e^v$ will be regarded as (possibly non--unitary) quantum gates acting on a single, respectively a pair, of $q$--level quantum systems. We now consider a 1D  quantum system composed of $n$ $q$--level systems and the quantum circuit ${\cal C}$ acting on this system as depicted in Fig.~\ref{fig:EM}. The circuit ${\cal C}$ consists of alternating layers of operations associated with the horizontal and vertical edges of the 2D lattice. Each round of ${\cal C}$ associated with a layer of horizontal edges consists of a product of one--local operators $W_e^h$, whereas every round associated with a layer of vertical edges is a product of (commuting) two--local operations $W_e^v$. 
We define computational basis states $|L\ra$ and $|R\ra$ associated to the left and right boundary conditions, respectively, analogously to Eq.~\eqref{eq:LR}. With these definitions, one has the following correspondence:
 \be
 \label{Z_with_boundary} 
 {\cal Z}_{\mbox{\scriptsize{em}}}^{L,R} = \langle L|{\cal C}|R\rangle\, .
 \ee 
This equation is readily verified by employing the definitions of gates $W_e^h$ and $W_e^v$.

\bfi[htb]\centering
\psfrag{a}{\small{$s_1^L$}}
\psfrag{b}{\small{$s_2^L$}}
\psfrag{c}{\small{$s_n^L$}}
\psfrag{d}{\small{$s_1^R$}}
\psfrag{e}{\small{$s_2^R$}}
\psfrag{f}{\small{$s_n$}}
\psfrag{i}{\small{$i$}}
\psfrag{j}{\small{$j$}}
\psfrag{k}{\small{$k$}}
\psfrag{l}{\small{$l$}}
\psfrag{w}{\small{$w_{ijkl}$}}
\psfrag{g}{\small{$\la s_1^L|$}}
\psfrag{h}{\small{$\la s_2^L|$}}
\psfrag{o}{\small{$\la s_n^L|$}}
\psfrag{u}{\small{$w_{ijkl}$}}
\psfrag{1}{\small{$i$}}
\psfrag{2}{\small{$j$}}
\psfrag{3}{\small{$k$}}
\psfrag{4}{\small{$l$}}
\psfrag{p}{\small{$| s_1^R\ra$}}
\psfrag{q}{\small{$| s_2^R\ra$}}
\psfrag{r}{\small{$| s_n^R\ra$}}
\psfrag{x}{\small{$s_1^L$}}
\psfrag{y}{\small{$s_2^L$}}
\psfrag{z}{\small{$s_3^L$}}
\psfrag{X}{\small{$s_1^R$}}
\psfrag{Y}{\small{$s_2^R$}}
\psfrag{Z}{\small{$s_3^R$}}
\psfrag{I}{\small{$i$}}
\psfrag{J}{\small{$j$}}
\psfrag{K}{\small{$k$}}
\psfrag{m}{\small{$w^h_{ij}$}}
\psfrag{n}{\small{$w^v_{jk}$}}
\psfrag{5}{\small{$i$}}
\psfrag{6}{\small{$j$}}
\psfrag{7}{\small{$k$}}
\psfrag{M}{\small{$W^h_{ij}$}}
\psfrag{N}{\small{$W^v_{jk}$}}
\psfrag{G}{\small{$\la s_1^L|$}}
\psfrag{H}{\small{$\la s_2^L|$}}
\psfrag{O}{\small{$\la s_n^L|$}}
\psfrag{P}{\small{$| s_1^R\ra$}}
\psfrag{Q}{\small{$| s_2^R\ra$}}
\psfrag{R}{\small{$| s_n^R\ra$}}
\psfrag{t}{\small{time}}
\hspace{-4mm}\includegraphics[width=1\columnwidth]{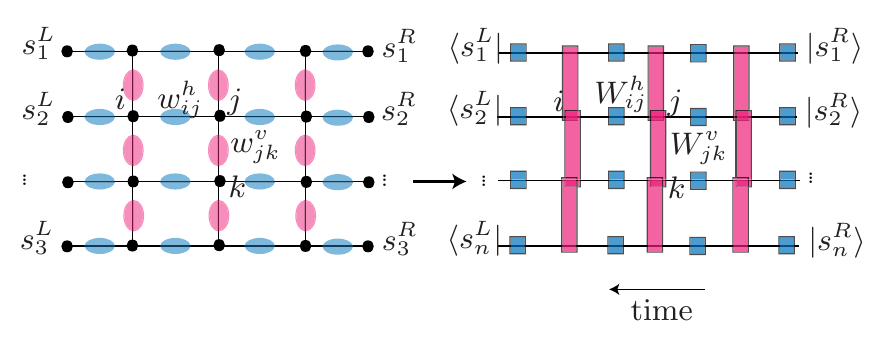}
\caption{
(Left) In an edge model, particles (black dots) sit at the vertices and interactions (pale red and blue ellipses) take place along the edges. (Right) This model is mapped to a quantum circuit, where particles are mapped to qubits, interactions along the time direction become single--qubit gates, and interactions perpendicular to time become diagonal two--qubit gates.
}
\label{fig:EM}
\efi

We emphasize that the comments concerning the mapping for vertex models apply to this mapping with straightforward modifications: unitary gates lead to complex coupling strengths (see Sec.~\ref{ssec:vertex-models} remark \eqref{im:vertex-complex}), and
the mapping can be extended to open and periodic boundary conditions (see Sec.~\ref{ssec:vertex-models} remark \eqref{im:vertex-OPBC}). 
Now we give some further modifications of the mapping for edge models that will be useful for the following sections. 
\ben[(i)]
\im
Consider that at site $i$ in the lattice a local magnetic field is present. This is represented by an additional term $h_i(s_i)$ in the energy function and corresponding Boltzmann weight 
\be
w_i(s_i)= e^{-\beta h_i(s_i)}\, ,
\ee
with $s_i=0, \dots, q-1$. We associate to it the following diagonal $q\times q$ matrix 
\be
\label{W_a} 
W_i:= \sum_{s_i} w_i(s_i)|s_i\rangle \langle s_i|\, .
\ee 
Now a mapping to a quantum circuit ${\cal C}$ can be established in a similar fashion as above, with the distinction that each layer associated with a slice of vertical edges now consists of a product of the associated two--qubit gates $W_e^v$ and the associated single--qubit gates $W_i$. Note that, as all such gates are diagonal operations, there is no problem regarding operator ordering. With this choice of ${\cal C}$, it can readily be verified that the associated partition function can be written as \eqref{Z_with_boundary}.

\im
These mappings may be easily generalized to graphs other than the 2D lattice, also similarly as for vertex models (see Sec.~\ref{ssec:vertex-models} remark \eqref{im:vertex-geometries}). In particular, below we will consider the following class of subgraphs of the 2D square lattice: a graph $G$ is said to be a \emph{planar circuit graph} if it can be obtained from an $n\times m$ rectangular grid (for some $n$ and $m$) by \emph{deleting} a subset of \emph{vertical} edges and \emph{contracting} a subset of \emph{horizontal} edges. We call $n$ the vertical dimension of $G$; note that this quantity is uniquely defined for every planar circuit graph. 
Similar as for the 2D square lattice, one can associate a quantum circuit ${\cal C}$ with every planar circuit graph; more precisely, such circuit acts $n$ $q$--level systems, and one associates each horizontal and vertical edge $e$ with the gates $W_e^h$ and $W_e^v$, respectively. Furthermore, local magnetic fields acting on the particles can also be easily incorporated by associating a gate $W_i$ with each vertex $i$; see Fig.~\ref{fig:planarcircuit} for an example. This mapping will be relevant in Sec.~\ref{ssec:Ising}.
\een

\bfi[htb]\centering
\psfrag{x}{\small{$s_1^L$}}
\psfrag{y}{\small{$s_2^L$}}
\psfrag{z}{\small{$s_3^L$}}
\psfrag{X}{\small{$s_1^R$}}
\psfrag{Y}{\small{$s_2^R$}}
\psfrag{Z}{\small{$s_3^R$}}
\psfrag{I}{\small{$w_i$}}
\psfrag{m}{\small{$w_{ij}$}}
\psfrag{n}{\small{$w_{jk}$}}
\psfrag{5}{\small{$W_{i}$}}
\psfrag{M}{\small{$W^h_{e}$}}
\psfrag{N}{\small{$W^v_{e}$}}
\psfrag{G}{\small{$\la s_1^L|$}}
\psfrag{H}{\small{$\la s_2^L|$}}
\psfrag{O}{\small{$\la s_n^L|$}}
\psfrag{P}{\small{$| s_1^R\ra$}}
\psfrag{Q}{\small{$| s_2^R\ra$}}
\psfrag{R}{\small{$| s_n^R\ra$}}
\psfrag{t}{\small{time}}
\hspace{-4mm}\includegraphics[width=1\columnwidth]{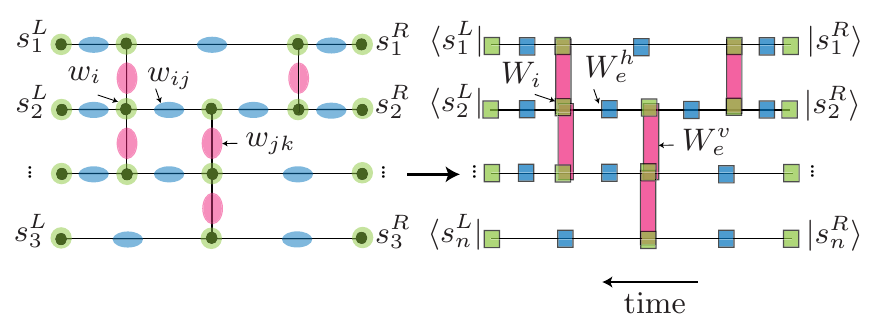}
\caption{(Left) Edge model with magnetic fields defined on a planar circuit graph. The latter is obtained from a rectangular grid by deleting some vertical edges and contracting some horizontal ones. (Right) This model is mapped to a quantum circuit: particles are mapped to qubits, and horizontal and vertical edge interactions are mapped to single--qubit (non--diagonal) and two--qubit diagonal gates, respectively, and local interactions (e.g.~magnetic fields) are mapped to single--qubit diagonal gates. 
}
\label{fig:planarcircuit}
\efi

In this paper we will be interested in two particular edge models: the Ising model and the Potts model. 
We now specialize the above discussion to the case of the 2D Ising model. We consider the Ising with magnetic fields defined on (sublattices of) a 2D square lattice. The interaction between spins $s_i$ and $s_j$ located at the endpoints of edge $e=(i,j)$ is given by
\be
h_e(s_i, s_j)=-J_e \delta(s_i + s_j)\, , 
\ee
and the contribution of the magnetic field at site $i$ is 
\be
h_i(s_i) = -h_i \delta(s_i)\, .
\ee 
Here the spin states $s_i$ may take values 0 and 1, the sums are performed modulo 2 and as before $\delta(0)=1$ and it is 0 otherwise. 
Further, $J_e$ and $h_i$ are constants which represent the strengths of the pairwise interaction and magnetic field, respectively. With the definitions \eqref{eq:W_e_horizontal}, \eqref{eq:W_e_vertical} and \eqref{W_a}, we have
\be
\ba{rcl} 
W_e^h&=& 
\left[ \begin{array}{cc} e^{\beta J_e}& 1\\ 1& e^{\beta J_e}\end{array}\right]\, , \quad
W_e^v =\mbox{ diag}(e^{\beta J_e}, 1, 1, e^{\beta J_e})\, ,\\ 
W_i&=& 
\left[ \begin{array}{cc} e^{\beta h_i}&0 \\ 0&1\end{array}\right]\, .
\ea
\lb{eq:WIsing}
\ee

Now we concentrate on the mapping for the Potts model. 
Given a graph $G$ with vertex set $V$ and edge set $E$, the Potts model \cite{Wu84} consists of $q$--level particles sitting at the vertices of $G$ and interacting along the edges of $G$. 
Let $u,v$ denote two such $q$--level particles $u,v\in\{0,1,\ldots,q-1\}$, which interact along edge $e$. Then the Potts--type interaction is of the form
\be
h^{e}(u,v) = -J_{e} \delta(u-v) \, .
\lb{eq:Potts}
\ee
where $\delta(0) = 1$ and it is 0 otherwise. 
We shall later consider a more general form of this interaction:
\be
h^{e}(u,v) = -J_{u=v} \delta(u-v) - J_{u\neq v} (1-\delta(u-v) )\, .
\lb{eq:Potts-our}
\ee
This amounts to a shift of the interaction energy which does not change the physics. Let $\mf{s}$ denote the spin state of all particles: $\mf{s}=(u,v,\ldots)$. Then the Hamiltonian of the Potts model is a sum of these two--local terms over all edges 
\be
H (\mf{s})= \sum_{e\in E} h^{e}(u,v)\, .
\lb{eq:HPotts}
\ee

\subsection{Lattice gauge theories}
\label{ssec:LGTs}

Now we focus on another family of  models, namely $\mathbb{Z}_2$ LGTs (see Sec.~\ref{ssec:background}), and we will introduce mappings for their partition functions. 

We shall restrict the following discussion to a 3D $\mathbb{Z}_2$ LGT with the  temporal gauge. As explained in Sec.~\ref{ssec:background}, fixing this gauge is achieved by fixing all spins lying on edges with a specific direction of the lattice (the `time' direction). 
However, the mapping can be generalized in a straightforward manner to 3D $\mathbb{Z}_2$ LGTs with another gauge fixing of the spins---we will return to this comment below and in Sec.~\ref{ssec:3DZ2LGT}. 
Because of the temporal gauge fixing, interactions in the time direction are \emph{two--body} interactions, whereas those in the spatial direction (i.e.~in faces without an edge in the time direction) remain four--body interactions as originally. 
To construct the mapping, we proceed similarly as above.
We associate the Boltzmann weight of the two--body interaction at the temporal face
\be
w_f^t (s_i,s_j) := e^{-\beta h_f(s_i,s_j)}\, ,
\ee
with a single--qubit (non--diagonal) gate $W_f^t$, 
\be
W_f^t : =\sum_{s_i,s_j} w_f^t (s_i,s_j) |s_j\ra \la s_i|\, .
\lb{eq:singlequbitgate}
\ee
Further, the Boltzmann weight of the four--body interaction at the spatial face
\bea
w_f^s (s_i,s_j,s_k,s_l):= e^{-\beta h_f(s_i,s_j,s_k,s_l)}
\label{eq:F}
\eea
is mapped to a four--qubit diagonal gate $W_f^s$
\bea
W_f^s &:=& \sum_{s_i,s_j,s_k,s_l} 
w_f^s (s_i,s_j,s_k,s_l)\times \nn\\
&&|s_i, s_j, s_k, s_l\ra \la s_i, s_j, s_k, s_l |\, .
\label{eq:F}
\eea
Thus, this maps the 3D $\mathbb{Z}_2$ LGT to a quantum circuit where a 2D array of qubits is processed in the time direction with single--qubit gates, and in the spatial direction with (diagonal) four--qubit gates (see Fig.~\ref{fig:LGT-mapping}). 
As before, it follows that the partition function $\mc{Z}=\sum_{\mf{s}} \prod_f w_f^s w_f^t$ is mapped to a quantum a circuit  $\mc{C} = \prod_f W_f^s W_f^t$.
Let $L,R$ denote the left and right boundaries, respectively, as in Eq.~\eqref{eq:LR}. Then our mapping reads
\be
{\cal Z}_{\mbox{\scriptsize{LGT}}}^{L,R} = \la L |{\cal C}|R\ra\, .
\lb{eq:ZLGT}
\ee

\bfid[htb]\centering
\psfrag{x}{\small{$s_1^L$}}
\psfrag{y}{\small{$s_2^L$}}
\psfrag{X}{\small{$s_1^R$}}
\psfrag{Y}{\small{$s_2^R$}}
\psfrag{u}{\small{$\la s_1^L|$}}
\psfrag{v}{\small{$\la s_2^L|$}}
\psfrag{U}{\small{$|s_1^R\ra $}}
\psfrag{V}{\small{$|s_2^R\ra $}}
\psfrag{o}{\small{$0$}}
\psfrag{t}{\small{time}}
\psfrag{w}{\small{$w_f^t (s_i,s_j)$}}
\psfrag{W}{\small{$W_{f}^t$}}
\psfrag{s}{\small{$w_s^t (s_i,s_j,s_k,s_l)$}}
\psfrag{S}{\small{$W_{f}^s$}}
\includegraphics[width=1.55\columnwidth]{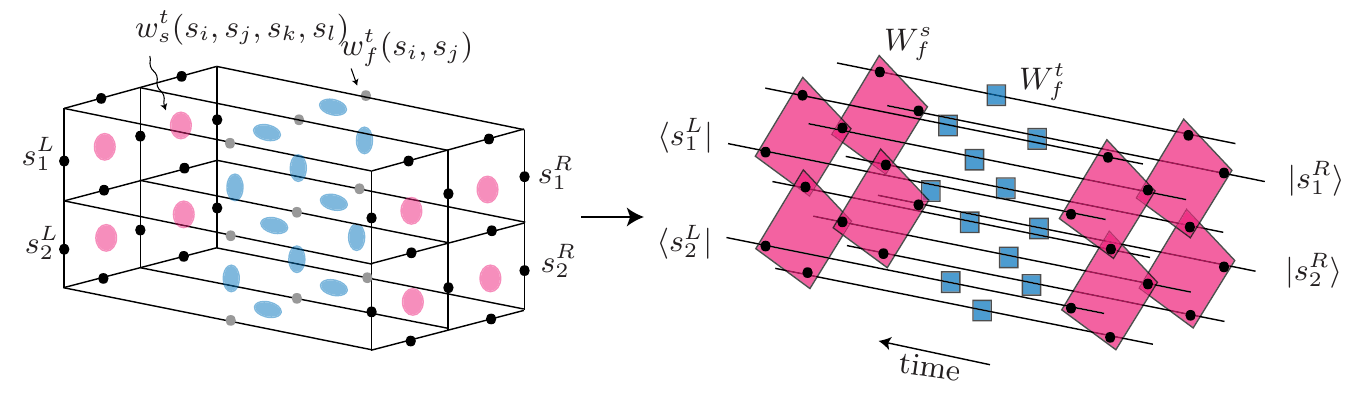}
\caption{
(Left) In a 3D LGT, particles (black and gray dots) sit at the edges and interactions (pale red and blue ellipses) take place on the faces. Gray dots indicate particles whose state has been fixed by the gauge. 
(Right) This model is mapped to a quantum circuit, where interactions along the time direction become single--qubit gates (blue squares), and those perpendicular to it become four--qubit gates (pale red squares). 
Thus, a 2D array of qubits is processed by a circuit consisting of a sequence of single--qubit gates and diagonal four--qubit gates.}
\label{fig:LGT-mapping}
\efid

Similarly as for vertex and edge models, we note that unitary gates will be translated to complex coupling strengths (as in Sec.~\ref{ssec:vertex-models} remark \eqref{im:vertex-complex}), and one can obtain similar mappings for LGTs with open and periodic boundary conditions (see Sec.~\ref{ssec:vertex-models} remark \eqref{im:vertex-OPBC}).
Next we make some further comments about this construction.
\ben[(i)]
\im
The mapping can easily be extended to $\mathbb{Z}_2$ LGTs with other gauge fixings. As a matter of fact, we will use one of these mappings in Sec.~\ref{ssec:3DZ2LGT}. Since one deals with square lattices, one can always associate one dimension of the lattice to time. Then, generally, one fixes some spins in the time direction and some others in the space direction, the only restriction being the avoidance of closed loops (see the comment in Sec.~\ref{ssec:background} and \cite{Cr77}).
This introduces new type of interactions. For example, there can be a face in the space direction where three spins have been fixed by the gauge, and thus only one variable is left:
\be
w_f (s_i)= e^{-\beta h_f(s_i)}\, .
\ee
Such interactions are mapped to single--qubit \emph{diagonal} gate $W_f^s$:
\bea
W_f^s &=& \sum_{s_i} w_s(s_i)|s_i\ra \la s_i|\, .
\label{eq:F}
\eea
Examples of these gates will be given in Sec.~\ref{ssec:3DZ2LGT}. 
Note, however, that the mapping does not need to be defined on all faces: for the \ts{BQP}--completeness proof it suffices to specify the coupling strength on every face. In particular, we will set $J=0$ on faces in  the time direction where the temporal gauge is \emph{not} fixed, and the interactions are not mapped to a gate in this case. 
\im
Notice that, due to the form of the interaction of this system (given by \eqref{eq:int-LGT}), each gate is specified by only one parameter: $e^{\beta J}$. We will make repeated use of this fact in Sec.~\ref{ssec:3DZ2LGT}.
\im
This mapping can be extended to $\mathbb{Z}_q$ LGTs (see~\cite{Ko79} for an introduction of these models). In this case, each interaction can take $q$ values, and they would translate to one-- and four--\emph{qudit} gates, where each qudit is a $q$--level quantum system.
\im
More generally, the mapping can also be extended to $\mathbb{Z}_2$ LGTs defined on square lattices in $d$ dimensions. 
The constructions is a generalization of the 3D case, where one has to select a translational invariant time direction out of the $d$ dimensions, and fix the temporal gauge.
Since interactions also take place on the faces, time and space interactions are also mapped to one-- and four--qubit gates, respectively.
This results in a quantum circuit that processes a $(d-1)$ dimensional array of spins with single--qubit gates in the time direction, and with four--qubit gates in the space dimensions. 
Similar considerations apply to $\mathbb{Z}_q$ LGTs, with one-- and four--qudit gates.
\een

\section{\ts{BQP}--completeness results}
\lb{sec:BQP}

This section contains the central results of this paper: 
here we will provide efficient quantum algorithms to estimate the partition function of the six vertex model, the Ising model, the Potts model and the 3D $\mathbb{Z}_2$ LGT in a certain (complex) parameter regime,
with polynomial accuracy.
Moreover, we will show that approximating these partition functions is \textsf{BQP}--complete. 

\subsection{Six vertex model}
\label{ssec:6VM}

Our goal is to prove that approximating the partition function ${\cal Z}$ of some  vertex models in a certain (complex) parameter regimes is \ts{BQP}--complete (see also \cite{Va09}). To show that, we will make use of the mappings between vertex models and quantum circuits described in Sec.~\ref{ssec:vertex-models}.

We consider the $q=2$ six vertex model (or: `ice--type model') and the eight vertex model \cite{Ba82} on a (tilted) 2D square lattice. In the six vertex model, only 6 of the 16 possible spin configurations give rise to a non--zero Boltzmann weight. More precisely, $W^{a}$ is a $4\times 4$ matrix of the form
\be
\label{8VM} 
W^{a} = \left(\begin{array}{cccc} w_{00,00} & 0 & 0&0\\ 0 &  w_{01,01} & w_{01,10} & 0\\ 0 &  w_{10,01} & w_{10,10} & 0\\0 & 0 & 0&w_{11,11}\end{array} \right)\, ,
\ee
where we use $w_{s_i,s_j,s_k,s_l}$ as a shorthand notation for $w(s_i,s_j,s_k,s_l)$ (see Eq.~\eqref{eq:Wa}). 
The eight vertex model \cite{Ba82} is obtained by additionally allowing the entries $w_{00,11}, w_{11,00}$ to be non--zero. We consider a parameter regime of the classical model where all matrices $W^a$ are unitary. This gives rise to a unitary circuit ${\cal C}$ formed of two--qubit quantum gates. Notice that this generally corresponds to (non--physical) complex parameters for either coupling strengths $J$ or the inverse temperature $\beta$ as we pointed out in Sec.~\ref{ssec:vertex-models} remark \eqref{im:vertex-complex}. Finally, we assume that we have staggered left an right boundary conditions of the form $L = R = (0101\dots)$. Our result is the following.

\begin{res}
\textnormal{\textbf{(\ts{BQP}--completeness of the six vertex model)}}
\lb{thm:BQP6VM}
Consider the six vertex model defined on a $n \times poly(n)$ rectangular grid 
with fixed boundary conditions.
 Further, consider that this is defined at inverse temperature $\beta$ and with couplings strengths
\be
\ba{rcl}
w_{00,00}&=&w_{11,11}=e^{i2t}\\ 
w_{01,01}&=&w_{10,10}=\cos(2t)\\
w_{01,10}&=&w_{10,01}=i\sin(2t)\, ,
\ea
\lb{eq:6VM-complex-1}
\ee
where $t$ is a continuous parameter, and
\be
\ba{rcl}
w_{00,00}&=&w_{11,11}=1\\
w_{01,01}&=&w_{10,10}=w_{01,10}=-w_{10,01}=\frac{1}{\sqrt{2}}\, .
\ea
\lb{eq:6VM-complex-2}
\ee
Let $\mc{Z}$ denote the partition function of this model. 
Then we provide efficient quantum algorithms to estimate 
$\mc{Z}$
with polynomial accuracy.
We also show that the problem of approximating 
$\mc{Z}$ is \ts{BQP}--complete.
\end{res}

Before starting the proof, we remark boldface symbols will denote encoded states and operators throughout this paper. To prove Result~\ref{thm:BQP6VM}, we will show that any quantum computation can be reduced to the evaluation of the partition function of a six vertex model on a tilted 2D square lattice with staggered boundary conditions.
We prove this statement in the following steps: 
\ben
\im
We show that quantum gates of the form \eqref{8VM} are computational universal for {\em encoded} quantum computation. To do so, we use the four--qubit encoding for $|\bs{0}\ra$ given by Refs. \cite{Di00,Hs03}, which is of the form:
\be
|\bs{0}\rangle = \frac{1}{2}(|01\rangle - |10\rangle)^{\otimes 2}\, .
\ee
Note that $|\bs{1}\rangle$ can be prepared by means of the encoded universal circuit that we will show next. 

Now we consider the exchange (or Heisenberg) interaction,
\be
H_{\mbox{\scriptsize{ex}}} =\sigma_x\otimes \sigma_x + \sigma_y\otimes \sigma_y + \sigma_z\otimes \sigma_z\, ,
\ee
with corresponding two--qubit gates
\be
\label{UHeis}
U=e^{it H_{\mbox{\scriptsize{ex}}}}\, .
\ee
The Heisenberg interaction is (encoded) universal for quantum computation \cite{Di00,Hs03}. 
In other words, by using gates of the form \eqref{UHeis}, one can prepare any quantum state $|\bs{\psi}\ra = \bs{{\cal C}}|{\bs 0}\rangle$ in an encoded form. 
Gates of the form \eqref{UHeis} can be generated with six vertex--type gates, i.e.~of the form~\eqref{8VM}, by setting the non--zero entries specified in \eqref{eq:6VM-complex-1}. 

\im
Now we show that the encoded initial state $|\bs{0}\rangle^{\otimes N}$ can be prepared from the state corresponding to staggered boundary conditions $|0101\ldots \rangle$. To this aim, we consider an operation $V$ of the form \eqref{8VM} with the non--zero entries of Eq.~\eqref{eq:6VM-complex-2}. 
It is straightforward to check that $V|01\rangle = (|01\rangle -|10\rangle)/\sqrt{2}$, and hence
\be
|\bs{0}\rangle =V^{\otimes 2} |0101\ra\, .
\ee
\im
Finally, we observe that matrix elements of the form 
$\la \bs{0} | ^{\bigotimes n} \bs{\mathcal{\cal C}}|\bs{0} \ra^{\bigotimes n}$ can be efficiently approximated by a quantum computer with polynomial accuracy, as long as $\bs{{\cal C}}$ can be implemented efficiently. Since we are dealing with a poly--size quantum circuit consisting of two--qubit gates (as indicated in Eq.~\eqref{8VM}),  $\bs{{\cal C}}$ can be implemented efficiently. 
The estimation of the matrix element is achieved by the Hadamard test, as pointed out in Sec.~\ref{sec:matrix_elements_BQP}. 
These overlaps are related to partition functions of the six vertex model in a certain (complex) parameter regime via \eqref{partitionVM}, since all gates $U,V$ involved are of the form \eqref{8VM}, and thus correspond to Boltzmann weights of the six vertex model. 
\een
This concludes the proof of Result \ref{thm:BQP6VM}.

We make a few remarks concerning this construction.
\ben[(i)]
\im \lb{im:real-positive}
Note that the complex parameter regime of Eqs.~\eqref{eq:6VM-complex-1} and \eqref{eq:6VM-complex-2}  is due to the fact that these entries correspond to unitary gates, as noted in Sec.~\ref{ssec:vertex-models} remark \eqref{im:vertex-complex}.

It is worth mentioning that in \cite{Sh03} it is shown that universal quantum computation can be achieved with real gates alone (i.e. gates with real entries). Thus, at first sight, it may seem that applying this method to our circuits would allow us to prove results for a real parameter regime of the classical spin models. However, we remark that the entries of our entries correspond to the Boltzmann weights of the interactions, which are not only real but also \emph{positive}. The latter condition is not satisfied in \cite{Sh03}. 

\im 
We observe that universality is already obtained for a suitable discrete set of unitary gates $e^{it H_{\mbox{\scriptsize{ex}}}}$ \cite{Di00,Hs03}, which leads to a discrete set of Boltzmann weights in the corresponding six vertex model.
\een

\subsection{Ising model}
\label{ssec:Ising}
Here we show that approximating the Ising partition function in the presence of an external magnetic field is a \ts{BQP}--complete problem. More precisely, we find the following.

\begin{res}
\textnormal{\textbf{(\ts{BQP}--completeness of the Ising model)}}
\label{thm_BQP_ising}
Consider any planar circuit graph $G$. Let $\tau$ denote the number of horizontal edges in $G$ and let $n$ be its vertical dimension. Consider a classical Ising model at inverse temperature $\beta$ defined on $G$, where on each site a constant (complex) magnetic field $h_a$ is present satisfying $e^{\beta h_a}=e^{\frac{i\pi}{4}}$, and on each edge a constant (complex) coupling $J_e$ is present satisfying $e^{\beta J_e}=i$.
Let ${\cal Z}$ denote the partition function of the model with open boundary conditions.  
Then we provide efficient quantum algorithms to estimate
\be
\lb{BQP_ising} 
\frac{{\cal Z}}{\kappa}, \quad\kappa:= 2^{\frac{\tau}{2} + n}\, , 
\ee 
with polynomial accuracy. 
We also show that the problem of estimating \eqref{BQP_ising} with this accuracy is \ts{BQP}--complete. 
\end{res}

The proof will consist of several steps. Gates corresponding to this model are of the form \eqref{eq:WIsing}; in particular, we consider the gates
\be
\ba{rcl}
W_h&:=& \left[ \begin{array}{cc}i& 1\\ 1& i\end{array}\right] \, ,\quad
W_v:= \trm{ diag}(i, 1, 1, i)\\
V&:=& \left[ \begin{array}{cc} e^{i\pi/4} &0 \\0 &1\end{array}\right]\, .
\ea
\ee

To show that \eqref{BQP_ising} can be approximated with polynomial accuracy in poly--time with a quantum computer, we use the mapping of an edge model in the presence of a local magnetic field (with open boundary conditions) to a quantum circuit ${\cal C}$ described in Sec.~\ref{ssec:edge-models}. Letting $n$ be the vertical dimension of $G$ as  in the statement of the result, the associated quantum circuit ${\cal C}$ is an $n$--qubit circuit composed of the gates $W_h, W_v$ and $V$. In particular, we have 
\be 
{\cal Z} = 2^n\langle +|^{\otimes n}{\cal C}|+\rangle^{\otimes n}\, ,
\ee 
where $|+\rangle = (|0\rangle + |1\rangle)/\sqrt{2}$. Note that, for the values of the magnetic fields and the couplings adopted in the statement of the result, the matrices $W_v$ and $V$ are unitary. Moreover, the matrix $ \bar W_h:= W_h/\sqrt{2}$ is unitary as well. Letting $\bar {\cal C}$ denote the unitary quantum circuit obtained by replacing every gate $W_h$ by $\bar W_h$, we simply have $\bar {\cal C} = {\cal C}/2^{\frac{\tau}{2}}$. It follows that 
\be 
\frac{{\cal Z}}{2^{\frac{\tau}{2} + n}} = \langle +|^{\otimes n}\bar {\cal C}|+\rangle^{\otimes n}\, ,
\ee 
where the right hand side now represents a matrix element of a poly--size, \emph{unitary} quantum circuit. From the Hadamard test it follows that estimating \eqref{BQP_ising} with polynomial accuracy is achievable in poly--time on a quantum computer.

Next we show that approximating \eqref{BQP_ising} with polynomial accuracy is \ts{BQP}--hard. This proceeds in the following steps:
\ben
\im
Denote $T:= V^{\frac{1}{2}} \bar W_h V^{\frac{1}{2}}$, where $V^{\frac{1}{2}}:= \trm{diag} (e^{\frac{i\pi}{8}},1)$. 
We will need the following result:
     
\begin{lem}[Universality of Ising--type gate set]
\label{univ-Ising-gate}
Any poly--size $n$--qubit quantum circuit composed of two--qubit unitary gates can be approximated to accuracy $\epsilon$ (with respect to the operator norm) by a circuit composed of poly$(n, \frac{1}{\epsilon})$ single--qubit gates $T$ and two--qubit gates $T^{\otimes 2} W_v T^{\otimes 2}$, where every such two--qubit gate is restricted to act on nearest--neighboring qubits only. Moreover the latter circuit can be found in time poly$(n, \frac{1}{\epsilon})$.
\end{lem}
This Lemma is proved in Appendix~\ref{app:univ-Ising-gate}; here we continue the argument to prove Result~\ref{thm_BQP_ising}.

\im  Consider an arbitrary poly--size $n$--qubit circuit $U$ composed of two--qubit unitary gates. It follows from the discussion in Sec.~\ref{sec:matrix_elements_BQP} that the problem of approximating general matrix elements $\langle+|^{\otimes n} U |+\rangle^{\otimes n}$ with polynomial accuracy is \ts{BQP}--hard. We now write $U$ as \be U=  [V^{\frac{1}{2}}]^{\otimes n}U'  [V^{\frac{1}{2}}]^{\otimes n}\ee for some suitable $U'$. Note that $U'$ is again a poly--size circuit, and can be found efficiently. We denote $A:= [V^{\frac{1}{2}}]^{\otimes n}$, i.e.~$U=AU'A$.

\im Due to Lemma \ref{univ-Ising-gate}, every poly--size circuit $U'$ can be approximated with accuracy $\epsilon$ by a circuit $U''$ of size poly$(n, 1/\epsilon)$ composed of $T$ gates and nearest--neighbor $T^{\otimes 2} W_v T^{\otimes 2}$ gates, and $U''$ can be found efficiently.
This implies that each matrix element $\langle+|^{\otimes n} U |+\rangle^{\otimes n}$, where $U$ is any poly--size circuit as before,  can be approximated with accuracy $1/$poly$(n)$ by a matrix element of the form  
\be 
\label{matrix_element_U'}\langle+|^{\otimes n} A U'' A|+\rangle^{\otimes n}\, , 
\ee  
where $U''$ is a poly--size circuit composed of $T$ gates and nearest--neighbor $T^{\otimes 2} W_v T^{\otimes 2}$ gates.  Hence, also the problem of approximating the matrix elements \eqref{matrix_element_U'} with 1/poly accuracy is \ts{BQP}--hard. Our goal is to show that any such matrix element coincides with a quantity ${\cal Z}/\kappa$ as considered in Result~\ref{thm_BQP_ising}.

\im By definition, the gates $T$ and $T^{\otimes 2} W_v T^{\otimes 2}$ contain square roots of the operation $V$. However, the operation $AU'' A$ only contains integral powers of the gate $V$. To see this, first note that any circuit $U''$ of $T$ and $T^{\otimes 2} W_v T^{\otimes 2}$ gates only contains $V^{\frac{1}{2}}$ gates at the left and right `boundary' of the circuit. In the circuit $A U'' A$, each $V^{\frac{1}{2}}$ gate at the boundary is multiplied with another $V^{\frac{1}{2}}$ gate due to the presence of the left and right boundary operator $A$. As a result, the operation $A U' A$ is a genuine circuit composed of $W_v$, $\bar W_h$ and $V$ gates. 

\im With any such circuit $A U'' A$ we now associate a graph $G$ in the following way. With each single--qubit gate $\bar W_h$ we associate a single horizontal edge, and with each two--qubit gate $W_v $ we associate a vertical edge. The entire graph $G$ associated with the quantum circuit is then obtained by simply `gluing' together these edges in the natural way. An example is given in Fig.~\ref{fig:planarcircuit}. Note that the graph is indeed a planar circuit graph, and that the vertical dimension of $G$ is $n$.
Moreover, consider an Ising model defined on $G$ with parameters $\beta$, $J_e$ and $h_e$ as in Result~\ref{thm_BQP_ising}, with open boundary conditions, and let ${\cal Z}$ denote the partition function of the model. Then ${\cal Z}$ can be expressed as a matrix element of the form \eqref{Z_free}, for some circuit ${\cal C}$ composed of $W_h$, $W_v$ and $V$ gates. It is now straightforward to verify that, up  to a normalization stemming from the presence of $1/\sqrt{2}$ in the gates $\bar W_h$, the circuit ${\cal C}$ coincides with the circuit $A U'' A$. More precisely, letting $\tau$ denote the number of horizontal edges in $G$, one has  ${\cal C} = 2^{\frac{\tau}{2}} AU'' A$. This shows that
\be
 \langle+|^{\otimes n} A U''A|+\rangle^{\otimes n}= \frac{{\cal Z}}{2^{\frac{\tau}{2} + n}}\, .
 \ee 
It follows that every matrix element of the form \eqref{matrix_element_U'} coincides with an Ising partition function with parameters $\beta$, $J_e$ and $h_a$ as in Result~\ref{thm_BQP_ising}, defined on the associated planar circuit graph $G$. This proves that the problem of estimating ${\cal Z}/{2^{\frac{\tau}{2} + n}}$ with polynomial accuracy is \ts{BQP}--hard.
\een

\subsection{Potts model}
\label{ssec:Potts}

In this section we show that approximating partition function of the three--level Potts model with complex parameters on a quasi 2D square lattice is \ts{BQP}--complete. To be precise, we find the following.

\begin{res}
\label{thm:BQPPotts}
\textnormal{\textbf{(\ts{BQP}--completeness of the Potts model)}}
Consider the Potts model with three--level particles
defined on a poly--size quasi 2D square lattice 
with fixed boundary conditions. 
Let $\beta$ denote the inverse temperature and $J_{u=v}$ ($J_{u\neq v}$) the coupling strength when the two interacting particles spins $u$ and $v$ are (not) in the same state (see \eqref{eq:Potts-our}). Consider this model defined in the following parameter regime 
\bea
(e^{\beta J_{u=v}},e^{\beta J_{u\neq v}}) &\in&  \{ 
(1,0),
(e^{i\pi/8},1),
(-i,1),\nn\\
&& 
(1,1),
(0,1),
(\epsilon,1),\nn\\
&& 
((\sqrt{2}\epsilon)^{-1},1),
(-1,1)
\}
\eea
where $\epsilon$ is a polynomially small number, i.e.~$\epsilon = \mc{O}(1/{\rm poly}(n))$. Moreover, consider that certain `blocks' of these coupling strengths appear together (that is, there are certain local distributions of couplings). 
Let  ${\cal Z}$ denote the partition function of this model.
Then we provide efficient quantum algorithms to estimate 
${\cal Z}$ with polynomial accuracy. 
We also show that estimating
${\cal Z}$ with this accuracy is \ts{BQP}--complete. 
\end{res}

To prove this result, we will make use of the idea of encoded universal quantum computation as in Sec.~\ref{ssec:6VM}. We will first present the encoding of physical qubits into logical ones, and then we will construct a universal gate set at the logical level with Potts--type gates. 

First of all we define an encoding. According to our mapping of Sec.~\ref{ssec:edge-models}, a three--level quantum particle (a \emph{qutrit}) is associated to each three--level classical particle.
Let $\{|0\ra, |1\ra, |2\ra\}$ denote the basis states of each qutrit. We use two of these qutrits (which we refer to as \emph{physical} qutrits) to encode one \emph{logical} qubit, whose states $|\bs{0}\ra$ and $|\bs{1}\ra$ are defined as 
\be
\ba{rcl}
|\bs{0}\ra &=& |0\ra  |1\ra \, , \\
|\bs{1}\ra& =& |1\ra  |2\ra \, .
\ea
\lb{eq:encoding-Potts}
\ee
We will refer to the first and the second physical qutrits as the `upper' and the `lower qutrits', respectively.
Note that, with this encoding, the input state $|\bs{0}\ra |\bs{0}\ra \ldots |\bs{0}\ra$ requires no preparation, since it can be fixed as an initial boundary condition, viz.~$|R\ra=|0\ra|1\ra \ldots |0\ra|1\ra$. 

\begin{figure}[htb]
\centering
\psfrag{2}{\small{$|2\ra$}}
\psfrag{a}{\small{$(1,1)$}}
\psfrag{c}{\small{$(e^{i\pi/8},1)$}}
\psfrag{i}{\small{$|\bs{\psi}\ra$}}
\psfrag{o}{\small{$\bs{I_1}|\bs{\psi}\ra$}}
\psfrag{I}{\small{$|\bs{\psi}\ra$}}
\psfrag{O}{\small{$\bs{P}_{\pi/8}|\bs{\psi}\ra$}}
\psfrag{A}{(a)}
\psfrag{B}{(b)}
\psfrag{t}{\small{time}}
\includegraphics[width=0.75\columnwidth]{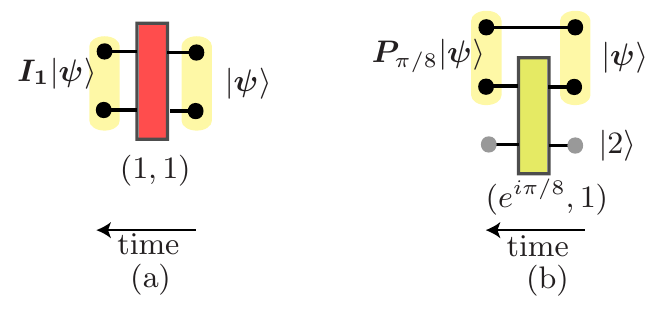}
\caption{
Auxiliary qubits (i.e.~fixed qubits) are depicted in gray, physical qubits in black, and logical qubits are highlighted in yellow.
(a) Logical single--qubit identity $\bs{I_1}$. (b) Logical phase gate $\bs{P}_{\pi/8}$. Each gate is determined by the pair of numbers $(\mu,\nu)$ (Eq.~\eqref{eq:ab}) which are indicated next to it, and its color is just a guide to the eye to identify equal gates.  Note that time runs in both figures from right to left.}
\label{fig:I1andP}
\end{figure}

Now we proceed with the construction of the encoded universal gate set.  
We first observe that, due to the Potts interaction \eqref{eq:Potts-our}, the gates \eqref{eq:W_e_horizontal} and \eqref{eq:W_e_vertical} contain only two different coefficients; for further reference, we denote them
\be
(\mu,\nu) := (e^{\beta J_{u=v}}, e^{\beta J_{u\neq v}})  \, .
\lb{eq:ab}
\ee
We construct the encoded universal gate set as follows:
\bi
\im The single--qubit identity
\be
\bs{I_1}=\sum_{\bs{i},\bs{j}=0}^1 \delta(\bs{u},\bs{v}) |\bs{u}\ra \la \bs{v}|\, ,
\ee
is achieved by applying a two--qutrit gate (between the two physical qutrits that compose the logical qubit) with $(\mu,\nu) =(1,1)$ (see Fig.~\ref{fig:I1andP}). 
\im
The phase gate 
\be
\bs{P}_{\pi/8} = |\bs{0}\ra\la \bs{0}| + e^{i\pi/8}|\bs{1}\ra\la \bs{1}|
\ee
is achieved by applying a two--qutrit gate between an auxiliary qutrit in the state $|2\ra$ and the lower qutrit of the logical qubit with $(\mu,\nu)= (e^{i\pi/8},1)$ (see Fig.~\ref{fig:I1andP}).

\im 
The Hadamard gate
\be
\bs{H}= \frac{1}{\sqrt{2}}\sum_{\bs{u},\bs{v}=0}^1 (-1)^{\bs{uv} } |\bs{v}\ra \la \bs{u} |\, 
\ee
is more involved and we refer the reader to Appendix \ref{app:HadamardPotts} for the details. 
\im
The two--qubit identity gate 
\be
\bs{I_2} &= &  \sum_{\bs{u},\bs{v}=0}^1 |\bs{uv}\ra \la \bs{uv}|\, ,
\ee
is trivially obtained by applying a two qutrit--gate between the two physical qutrits of the first logical qubit with $(\mu,\nu) = (1,1)$, and doing the same for the other logical qubit (see Fig.~\ref{fig:I2andCZ}).
\im
Finally, the phase gate between logical qubits 
\be
\bs{CZ} =   \sum_{\bs{u},\bs{v}=0}^1  (-1)^{\bs{uv}} |\bs{uv}\ra \la \bs{uv}|
\ee
is achieved by applying a two qutrit--gate between the lower qutrit of the first logical qubit and the upper qutrit of the second logical qudit with $(\mu,\nu)=(-1,1)$. We also need to apply the same gate between the lower qudit of the second logical qubit and an auxiliary particle in the state $|2\ra$ (see Fig.~\ref{fig:I2andCZ}).
\ei

\bfi[htb]
\centering
\psfrag{2}{\small{$|2\ra$}}
\psfrag{a}{\small{$(1,1)$}}
\psfrag{c}{\small{$(1,1)$}}
\psfrag{e}{\small{$(-1,1)$}}
\psfrag{g}{\small{$(-1,1)$}}
\psfrag{x}{\small{$|\bs{\psi_1}\ra$}}
\psfrag{y}{\small{$|\bs{\psi_2}\ra$}}
\psfrag{X}{\small{$|\bs{\psi_1}\ra$}}
\psfrag{Y}{\small{$|\bs{\psi_2}\ra$}}
\psfrag{o}{\small{$\bs{I_2} |\bs{\psi_1}\ra|\bs{\psi_2}\ra$ }}
\psfrag{I}{\small{$|\bs{\psi_1}\ra|\bs{\psi_2}\ra$}}
\psfrag{O}{\hspace{-5mm}\small{$\bs{CZ} |\bs{\psi_1}\ra|\bs{\psi_2}\ra$}}
\psfrag{A}{(a)}
\psfrag{B}{(b)}
\psfrag{t}{\small{time}}
\includegraphics[width=0.95\columnwidth]{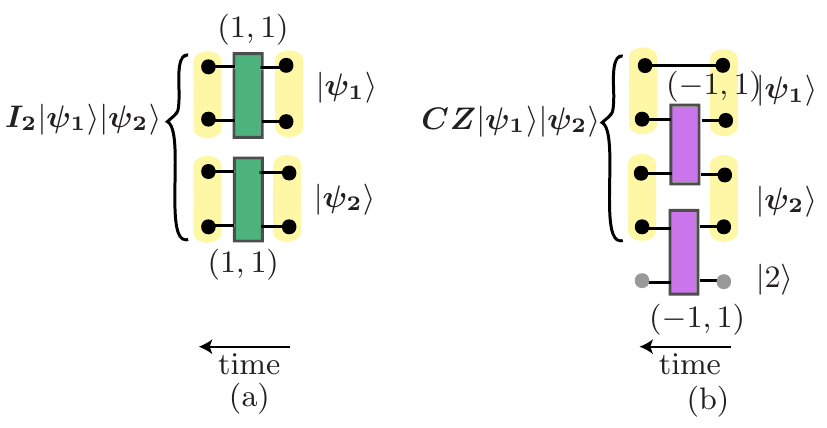}
\caption{
(a) Logical two--qubit identity $\bs{I_2}$. (b) Logical controlled phase gate $\bs{CZ}$. See the caption of Fig.~\ref{fig:I1andP} for the meaning of the pair of numbers and the color associated to each gate.}
\label{fig:I2andCZ}
\efi

In the construction presented above we need a constant supply of auxiliary qutrits, for which we envisage the following setup. We arrange the particles so that each physical qubit is in contact with four auxiliary qubits, as indicated in Fig.~\ref{fig:auxiliarytriangular}.
Moreover, each physical qubit is propagated in time via a single--qubit gate at the physical level (that is, a gate of the form \eqref{eq:W_e_horizontal}) which is achieved with $(\mu,\nu)=(1,0)$. 
 In this way, the physical qubits are distributed on a 2D square lattice, which is surrounded (in front and on the back) by auxiliary qubits---we refer to this as a quasi 2D square lattice. 

\bfi
[htb]
\centering
\psfrag{a}{\small{$|0\ra$}}
\psfrag{b}{\small{$|1\ra$}}
\psfrag{c}{\small{$|2\ra$}}
\psfrag{d}{\small{$|2\ra$}}
\psfrag{e}{\small{$|0\ra$}}
\psfrag{f}{\small{$|1\ra$}}
\psfrag{g}{\small{aux. }}
\psfrag{h}{\small{phys.}}
\psfrag{i}{\small{aux. }}
\psfrag{t}{\small{time}}
\psfrag{P}{\small{$|\bs{\psi}_{1}^t\ra$}}
\psfrag{T}{\small{$|\bs{\psi}_{2}^t\ra$}}
\psfrag{U}{\small{$|\bs{\psi}_{1}^{t+1}\ra$}}
\psfrag{V}{\small{$|\bs{\psi}_{2}^{t+1}\ra$}}
\psfrag{R}{\small{$|\bs{\psi}_{1}^t\ra$}}
\psfrag{S}{\small{$|\bs{\psi}_{2}^t\ra$}}
\psfrag{1}{\small{$(1,1)$}}
\psfrag{2}{\small{$(1,0)$}}
\psfrag{=}{\small{$=$}}
\psfrag{A}{(a)}
\psfrag{B}{(b)}
\includegraphics[width=0.95\columnwidth]{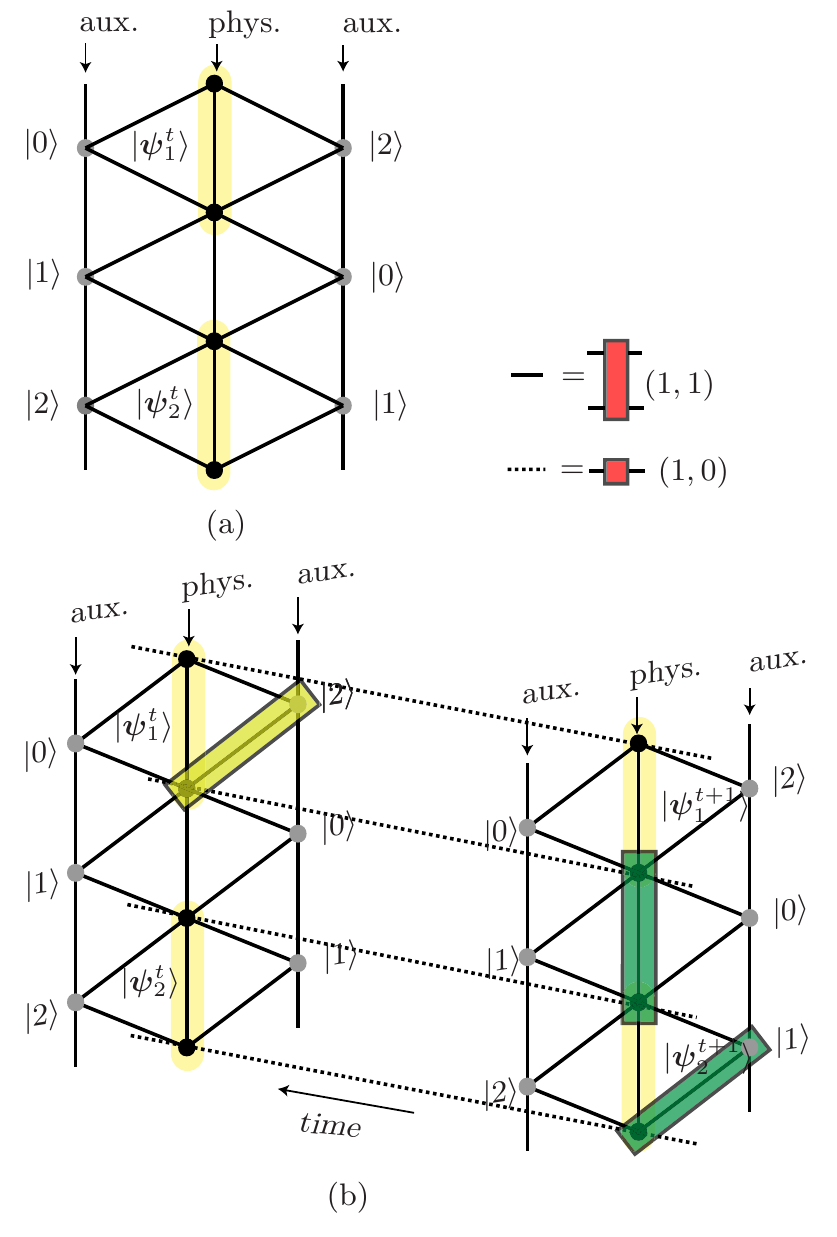}
\caption{
Setup for a Potts model with the physical qutrits surrounded by auxiliary qutrits (i.e.~qutrits whose value is fixed). 
Single--qutrit gates are applied in the time direction, and two--qutrit gates are applied in a fixed time slice.
Solid and dashed lines stand for the two--qubit and single--qubit identities (at the physical level), respectively, as specified in the legend.
(a) A time slice of the lattice at time $t$, where one can see the structure with two auxiliary qutrits connected to each physical qutrit. Each logical qutrit is composed of two  physical qutrits. (b) Two time slices of the lattice, at times $t$ and $t+1$. 
The figure shows an example of a circuit, where a phase gate $\bs{P}_{\pi/8}$ is applied to $|\bs{\psi}_1^t\ra$ and the gate $\bs{CZ}$ is applied to $|\bs{\psi}_1^{t+1}\ra|\bs{\psi}_2^{t+1}\ra$.
}
\label{fig:auxiliarytriangular}
\efi

This proves that we can perform encoded universal quantum computation using qutrits interacting with Potts--type nearest--neighbor gates. 
From the discussion of Sec.~\ref{sec:matrix_elements_BQP} it follows that estimating $\la \bs{0}|^{\otimes n} \bs{\mc{C}} |\bs{0}\ra^{\otimes n} = \mc{Z}$ with polynomial accuracy is \ts{BQP}--complete, where $\mc{Z}$ is the partition function specified in the statement of Result \ref{thm:BQPPotts}. This thus concludes the proof of Result~\ref{thm:BQPPotts}.

We point out that our \ts{BQP}--completeness result for the Potts model imposes constraints in the parameter regime as noted in Sec.~\ref{ssec:edge-models}. We now give some further remarks regarding this proof.
\ben[(i)]
\im 
\lb{im:Potts-remark-constraints}
Note that to realize a gate in more than one step (like the Hadamard gate, see Appendix~\ref{app:HadamardPotts}) implies that one must fix the value of all those coupling strengths \emph{together} in order to realize that gate in the quantum circuit. That is, the gate is only realized (and thus the complexity result is only achieved) if the classical model contains that `block' of coupling strengths. These are thus constraints in the distribution of couplings imposed by the realization of gates. 
\im
\lb{im:Potts-remark-infinity}
We observe that the single--qubit identity at the physical level $I_1$ requires to set $\nu=0$, which formally corresponds to setting $J_{u\neq v}=-\infty$. 
This unphysical regime can be avoided by letting $e^{\beta J_{u\neq v}}$ be a polynomially small number, that is  $e^{\beta J_{u\neq v}} = \mc{O}(1/\trm{poly}(n))$
By the same argument as for the Hadamard gate (see Appendix~\ref{app:HadamardPotts}), 
the resulting state will be polynomially close to the ideal one, and thus the error within the accuracy of the computation.

\im
We see with our construction that the Potts model in a 1D array is not likely to be \ts{BQP}--complete, whereas this model on a 2D setup (and with complex parameters) is \ts{BQP}--complete.
This is in agreement with the fact that the Potts model in 1D array or a tree--like structure is efficiently simulable classically \cite{Hu10}. 

\im
Finally, this result can be generalized to Potts model with any $q$. In this case, the encoded states are still given by Eq.~\eqref{eq:encoding-Potts}, and all gates are performed with the same procedure except for the Hadamard. For this gate, one adds filters for the $|3\ra\ldots |q-1\ra$ components on the upper and the lower qudit right after the filters for the $|2\ra$ and the $|0\ra$ component of Fig.~\ref{fig:hadamard-Potts}. That is, the number of filters scales linearly with $q$ (since one requires $q-2$ filters for each qudit). Moreover, each physical qudit has to be connected to $2 \lceil q/2\rceil $ auxiliary particles, each fixed in a different state, namely $|0\ra, \ldots, |q-1\ra$. This amounts to a similar construction to that of Fig.~\ref{fig:auxiliarytriangular} but where each physical qudit is connected to $\lceil q/2\rceil$ auxiliary qudits to the left and to the right. 

\een

\subsection{$\mathbb{Z}_2$ Lattice gauge theory}
\label{ssec:3DZ2LGT}

Now we turn to a different class of models, namely LGTs. Using the tools presented in Sec.~\ref{ssec:LGTs} we will prove the following result.

\begin{res}
\lb{thm:BQP3DZ2LGT}
\textnormal{\textbf{(\ts{BQP}--completeness of the 3D $\mathbb{Z}_2$ LGT)}}
Consider a $\mathbb{Z}_2$ lattice gauge theory defined on a 3D rectangular lattice of size $(4x,12y,7z)$, 
where $x,y$ and $z$ are natural numbers, and with fixed boundary conditions. 
 Let $\beta$ denote the inverse temperature and $J_f$ the coupling strength on its faces. Consider this model in the following complex parameter regime,
\be
e^{\beta J_f} \in \left\{
0,e^{i\xi}, \frac{1}{2}, \zeta\right\}\, ,
\ee
where $\xi$ is a continuous parameter $\xi \in [0,2\pi)$, and $\zeta$ is a polynomially small number,Ê
$\zeta = \mc{O}(1/\trm{poly}(n))$. 
Moreover, consider that certain `blocks' of couplings appear together, and 
that there is a certain gauge fixing.
Let ${\cal Z}$ denote the partition function of this model.
Then we provide efficient quantum algorithms to estimate 
\be
\frac{\cal Z }{\kappa}\, , \quad \kappa := 2^{\frac{xyz}{2}}
\lb{eq:Z-LGT}
\ee
with polynomial accuracy.
We also show that the problem of approximating \eqref{eq:Z-LGT} is \ts{BQP}--complete.
\end{res}

The proof of this result will proceed similarly as in the previous sections: we will construct an encoded universal gate set. We first define the following encoding of four physical qubits into one logical qubit:
\be
\ba{lll}
|\bs{0}\ra &=& |0\ra |0\ra |0\ra |0\ra \\
|\bs{1}\ra &=& |1\ra |1\ra |1\ra |1\ra \, .
\ea
\label{eq:encoding-LGT}
\ee
Note that preparing the input state $|\bs{0}\ra$ is trivially achieved by fixing the physical qubits (which correspond to physical spins) in $|0\ra |0\ra |0\ra |0\ra $. 
In the following we show how to construct a general single--qubit unitary gate (at the logical level), and a specific two--qubit gate (also at the logical level), namely (the non--local part of) a controlled phase gate.
\bi
\im 
First we express an arbitrary single--qubit rotation in its Euler decomposition,
\be
\bs{U}(\gamma,\beta,\alpha)=\bs{R_z}(\gamma ) \bs{H} \bs{R_z}(\beta ) \bs{H} \bs{R_z}(\alpha ) \, .
\lb{eq:Ualphabetagamma}
\ee
Thus, our goal is to generate arbitrary $z$--rotations and the Hadamard gate. 
The gate $\bs{R_z}(\xi)$ can be implemented by letting the logical qubit interact with a neighboring, auxiliary face which has all remaining qubits fixed to $|0\ra$, and setting $e^{\beta J} = e^{i\xi}$ in this face (see Fig.~\ref{fig:RzandCZ}). 

The logical Hadamard gate $\bs{H}$ is implemented as a teleportation--based gate. We refer the reader to Appendix \ref{app:HadamardLGT} for the details. 
We shall in fact apply a non--normalized version of this, namely 
$\tilde{\bs{H}} := \sqrt{2} \bs{H}$ (see below for a discussion of this normalization factor).

\im 
To generate the logical controlled phase gate $\bs{CZ}$ we decompose it into single--qubit $\bs{R_z}$ rotations and a two--qubit gate:
\bea
\bs{CZ} 
= \bs{R_z}^{(1)}(-\frac{\pi}{2})\bs{R_z}^{(2)}(-\frac{\pi}{2}) \:
\diag(1,i,i,1)\, .
\eea
The two--qubit gate $\diag(1,i,i,1)$ is implemented by letting the last two logical qubits interact via an auxiliary plaquette with $e^{\beta J}=i$ (see Fig.~\ref{fig:RzandCZ}). 
\ei

\bfi[htb]
\centering
\psfrag{a}{}
\psfrag{b}{}
\psfrag{c}{}
\psfrag{d}{}
\psfrag{e}{$|0\ra$}
\psfrag{f}{\hspace{-2mm}$|0\ra$}
\psfrag{g}{$|0\ra$}
\psfrag{l}{$e^{\beta J}\! =\!  0 $}
\psfrag{J}{$e^{\beta J}\! =\! e^{i\xi}$}
\psfrag{P}{$|\bs{\psi}\ra$}
\psfrag{T}{$|\bs{\psi}\ra$}
\psfrag{c}{$e^{\beta J} \! =\! i$}
\psfrag{K}{$|\bs{\psi}_1\ra$}
\psfrag{L}{$|\bs{\psi}_2\ra$}
\psfrag{r}{$|0\ra$}
\psfrag{s}{$|0\ra$}
\psfrag{A}{(a)}
\psfrag{B}{(b)}
\psfrag{C}{(c)}
\includegraphics[width=0.6\columnwidth]{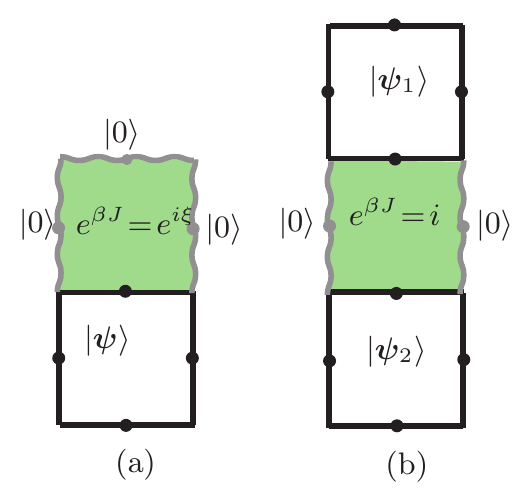}
\caption{The physical qubits composing the logical qubit are marked with thick, black lines, and wavy lines indicate qubits fixed by the gauge symmetry. Each gate is determined by $e^{\beta J}$.
(a) The logical rotation around the $z$--axis, $\bs{R_z}(\xi)$ is obtained by letting the logical qubit interact with an auxiliary plaquette whose spins are all fixed to $|0\ra$, and setting $e^{\beta J}=e^{i\xi}$ in this auxiliary plaquette. 
(b) The non--local part of $\bs{CZ}$, $\diag(1,i,i,1)$, is implemented by letting the two logical qubits interact via an auxiliary plaquette with $e^{\beta J} = i$.
 }
\lb{fig:RzandCZ}
\efi

Finally, we apply single--qubit identities at the physical level $I_1:=|0\ra\la 0| + |1\ra\la 1|$ on all physical qubits which belong to a logical qubit, unless otherwise specified. This gate is achieved by rescaling the interaction from $e^{\beta J_f}$ to $e^{\beta (J_f-1)}$, and then letting $e^{\beta (J_f-1)}$ tend to 0 polynomially fast (see Appendix~\ref{app:HadamardLGT}). 
We also set $J=0$ on the faces whose interaction is not mapped to a quantum gate, which sets their corresponding Boltzmann weights to 1, thus not `contributing' to the partition function. 
This is in accordance with the fact that these particles do not take part in the computation described by the quantum circuit.

Our construction of a circuit at the logical level differs from the general mapping presented in Sec.~\ref{ssec:LGTs} in the following sense. The quantum circuit processes a \emph{1D} array of logical qubits which are distributed along the $x$ direction. 
These are transformed by four--qubit physical gates in the $x$ direction (leading to logical single-- and two--qubit gates), and by single--qubit physical gates in the $z$ direction (which corresponds to time). The additional dimension ($y$ direction) is required for the realization of the teleportation--based Hadamard gates, which transport the 1D array that stores the quantum information in the $y$ direction.

Thus, we have constructed an encoded universal quantum circuit $\tilde{\bs{\mc{C}}}$ with $\mathbb{Z}_2$--LGT--type gates, where $\tilde{\bs{\mc{C}}}$ contains the non--normalized version of the Hadamard gates $\tilde{\bs{H}}$. 
Let $\bs{\mc{C}}$ denote the unitary circuit containing the normalized Hadamard gates $\bs{H} =\tilde{ \bs{H}}/\sqrt{2}$. Then we have that 
\be
\frac{\mc{Z}}{2^{xyz/2}} =  \la \bs{0}|^{\otimes n}\bs{\mc{C}} |\bs{0}\ra^{\otimes n}\, ,
\lb{eq:LGTZ=0C0}
\ee
where $xyz$ is the number of times that the gate $\tilde{ \bs{H}}$ is applied. 
From \eqref{eq:LGTZ=0C0} and the discussion of  Sec.~\ref{sec:matrix_elements_BQP} it follows that the estimating the partition function \eqref{eq:Z-LGT} with polynomial accuracy is \ts{BQP}--complete. 
This concludes the proof of Result~\ref{thm:BQP3DZ2LGT}.

We emphasize that, as in the previous results, we require complex parameters due to the unitary gates. 
Also the remark \eqref{im:Potts-remark-constraints} of Sec.~\ref{ssec:Potts} applies here: performing a logical gate in several steps (like the Hadamard gate, see Appendix \ref{app:HadamardLGT}) implies that a certain distribution of couplings must appear together in the classical model in order to prove our complexity result. 
Next we give some more comments concerning this result.
\ben[(i)]
\im 
Note that the single-- and two--qubit identity gates $\bs{I_1}$ and $\bs{I_2}$ can be generated using the universal gate presented above. For example, $\bs{R_z}(0)\bs{H}\bs{R_z}(0) \bs{H} \bs{R_z}(0) = \bs{I_1}$ and $ \bs{I_1}^{(1)}\bs{I_1}^{(2)} = \bs{I_2}$.
\im
Result~\ref{thm:BQP3DZ2LGT} is in contrast with the 2D $\mathbb{Z}_2$ LGT, whose complexity is `trivial', i.e.~it is in \ts{P} when the temporal gauge is fixed. 
On the other hand, the 3D $\mathbb{Z}_2$ LGT can be mapped via a duality transformation to the 3D Ising model \cite{Ko79}. Thus, they share the same complexity in the parameter regime given by this transformation.
We also observe that it was recently shown that computing the partition function of the 4D $\mathbb{Z}_2$ LGT in a real parameter regime is $\#$\ts{P}--complete \cite{De09b,De10}. 
Concerning the quantum computational complexity of the 4D $\mathbb{Z}_2$ LGT (and, more generally, of any $d$ $\mathbb{Z}_2$ LGT with $d>3$), one can make the following trivial observation: 
any of these models can be ``reduced'' 
(i.e.~become effectively equivalent) to the 3D $\mathbb{Z}_2$ LGT if 
the couplings in every face except those in a 3D volume are set to 0.
Thus, our results imply that their complexity in this (trivial) parameter regime is \ts{BQP}--complete. 
However, to the best of our knowledge, no quantum computational results are known outside this trivial parameter regime. 
\een

\section{Further results: one clean qubit model}
\lb{sec:further}

Here we point out a connection between the results obtained on the Ising model in Sec.~\ref{ssec:Ising} and a scheme for quantum computation called the `one clean qubit model'~\cite{Kn98}. In the latter, one considers a quantum computation where all qubits but the first one are initialized in the totally mixed state (and the first qubit is, say, in the state $|0\rangle$). To this initial state, an arbitrary poly--size quantum circuit may be applied, followed by a single--qubit measurement in the final stage of the computation. This one--clean--qubit model comprises a scheme that is believed to be weaker than the full power of quantum computers but stronger than classical computation (although these are unproved assertions). The corresponding complexity class of decision problems that can be solved efficiently with the one--clean--qubit scheme is called \ts{DQC1}.

A standard problem that can be solved using the one clean qubit model is the problem of estimating normalized traces of unitary quantum circuits. Let $U$ denote a poly--size $n$--qubit quantum circuit composed of, say, two--qubit gates. Then there exists an efficient quantum algorithm within the one clean qubit scheme which returns a number $c$ that provides (with exponentially small probability of failure) an $\epsilon$--approximation of the normalized trace $2^{-n} \mbox{Tr}(U)$ in poly$(n)$ time, for every $\epsilon$ that scales at most inverse polynomially with $n$. The technique is a simple variant of the Hadamard test and we refer to the literature \cite{Kn98}.
Moreover, the problem of estimating normalized traces of unitary quantum circuits is known to be \ts{DQC1}--hard. In other words, this problem captures all problems that can be solved efficiently within the one clean qubit paradigm. It thus plays a similar role as the unitary matrix element problem for \ts{BQP}.

Our results obtained in Sec.~\ref{ssec:Ising} immediately lead to a complete problem for \ts{DQC1} involving Ising partition functions defined on graphs with \emph{periodic boundary conditions}. We limit ourselves to a sketch of the argument. It follows from the discussion in Sec.~\ref{ssec:Ising} that, for every poly--size quantum circuit $U$ there exists an Ising model defined on a planar circuit graph $G$ (which can be found efficiently), with couplings and temperature as in Result~\ref{thm_BQP_ising} and with open boundary conditions, such that ${\cal Z}/\kappa$ provides a 1/poly approximation of the matrix element $\langle+|^{\otimes n}U|+\rangle^{\otimes n}$. Now,  instead of $|+\rangle^{\otimes n}$ we consider the matrix element $\langle s|U|s\rangle$ where $|s\rangle = |s_1\dots s_n\rangle$ represents an arbitrary computational basis state. This matrix element now coincides with ${\cal Z}^s/\kappa$, where ${\cal Z}^s$ denotes the partition function of the same model (i.e.~same graph, couplings and temperature), however considering boundary conditions where the left and right boundary spins are fixed in the same configuration $s$. Such a situation is equivalent to considering a graph $G'$ where each left boundary spin at the $k$th `row' in the graph is \emph{identified} with its corresponding right boundary spin, and the resulting spin is fixed in the state $s_k$, for all $k$. One thus arrives at an Ising model on a new graph $G'$ which is obtained from $G$ by enforcing \emph{periodic boundary conditions}, and where one vertical slice of spins is fixed in the configuration $s$. Finally, consider the normalized trace of the matrix element $U$, given by $2^{-n}\sum_s \langle s|U|s\rangle$. Due to the above discussion, this normalized trace may be approximated with 1/poly accuracy by  
\be 
\frac{1}{2^n\kappa}\sum_s {\cal Z}^s\ .
\ee
But as ${\cal Z}^s$ is the partition function on $G'$ with one vertical slice of spins fixed in the configuration $x$, the sum $\sum_s {\cal Z}^s \equiv {\cal Z}'$ simply represents the partition function on $G'$ where these spins are now fully summed out. The quantity ${\cal Z}'$ is hence the full--fledged partition function on $G'$ of the Ising model with couplings and temperature as before, and without any fixed spins or boundary conditions.

We thus arrive at the following result: consider any planar circuit graph $G$. Let $\tau$ denote the number of horizontal edges in $G$ and let $n$ be its vertical dimension. Let $G'$ be the graph obtained by enforcing periodic boundary conditions on $G$ (as above). Consider a classical Ising model at inverse temperature $\beta$ defined on $G$, where on each site a constant (complex) magnetic field $h_a$ is present satisfying $e^{\beta h_a}=e^{\frac{i\pi}{4}}$, and on each edge a constant (complex) coupling $J_e$ is present satisfying $e^{\beta J_e}=i$. 
Let ${\cal Z}'$ denote the partition function of the model.  Then the problem of approximating 
\be
\label{BQP_ising-DQC1} 
\frac{{\cal Z}'}{2^n\kappa}, \quad\kappa:= 2^{\frac{\tau}{2} + n}\, , 
\ee 
with polynomial accuracy, is \ts{DQC1}--hard.

\section{Conclusions}	
\lb{sec:conclusions}

In this work we have shown that estimating the partition function of the six vertex model on a 2D square lattice,
the 2D Ising model with magnetic fields on a planar graph,
the Potts model on a quasi 2D square lattice,
and the 3D $\mathbb{Z}_2$ LGT is \ts{BQP}--complete.  
All these models must be defined in a (partially) complex parameter regime in order to prove the result. 
In all but the Ising model, we further require that certain blocks of coupling strengths appear together.
Roughly speaking, that these problems are \ts{BQP}--complete means that they are as hard as simulating arbitrary quantum computation. 
Because our proof is constructive, we have also provided the efficient quantum algorithms that estimate the partition functions of the above models. 
The proofs are based on a mapping from partition functions to quantum circuits introduced in \cite{Va09} and extended here to the standard Ising model, Potts model and to $\mathbb{Z}_2$ LGTs. 
In this sense, our work puts these different kinds of classical spin model on an equal footing as far as their quantum computational complexity is concerned. 

It would be interesting to obtain quantum algorithms for computing such partition functions in a real parameter regime. At present we do not know whether our approach can be extended to prove such results. As we pointed in Sec.~\ref{ssec:6VM} remark \eqref{im:real-positive}, the Boltzmann weights give rise not only to real but also to positive entries in the quantum gates. The latter seems to be a rather severe restriction in the pursuit of this goal.

\section*{Acknowledgements}

We thank H.~J.~Briegel and J.~I.~Cirac for helpful discussions. 
This work was supported by the FWF and the European Union (QICS, SCALA, NAMEQUAM). 
MVDN acknowledges support by the excellence cluster MAP. 
MAMD thanks the Spanish MICINN grant FIS2009-10061, CAM research consortium QUITEMAD S2009-ESP-1594, European FET-7 grant PICC, UCM-BS grant GICC-910758. 

\appendix
\section{Universality of the Ising--type gate set}
\lb{app:univ-Ising-gate}

The proof of Lemma \ref{univ-Ising-gate} will use the following lemma.
\begin{lem}
\label{lem_ising}
Let $\epsilon>0$ and define $K:=\bar W_h V$. Up to a global phase, any two--qubit unitary operation can be approximated with accuracy $\epsilon$ (with respect to the operator norm) by a product of gates $K$ and $W_v$ i.e.~these gates form a strongly universal gate set.
\end{lem}
{\it Proof: } Let $P:=\trm{diag}(1, i)$ denote the phase gate, $H$ the usual Hadamard gate, $H=\sum_{i,j=0,1}(-1)^{ij} |i\ra\la j|$,  and $Z$ denote the standard Pauli $\sigma_z$ gate. We will also denote the controlled phase gate CZ $:=$ diag$(1, 1, 1, -1)$. Then the following identities can easily be verified: 
\be
\label{gate_identities} 
\ba{rcl}
ZKZKZ(K^{\dagger})^2Z &\propto& H \\
K ZK^{\dagger} ZK^{\dagger}ZK &\propto& P\, . 
\ea
\ee 
Now note that $W_v^2 \propto Z\otimes Z$. Thus Eqs.~\eqref{gate_identities} imply that 
\be
\label{gate_identities2} 
\ba{rcl}
W_v^2K_1W_v^2K_1 W_v^2[K_1^{\dagger}]^2 W_v^2 &\propto& H\otimes I \\
W_v^2 K_1^{\dagger} W_v^2 K_1 W_v^2 K_1 W_v^2K_1^{\dagger} &\propto&  P \otimes I\, .
\ea
\ee
So far we have showed that the operations $H\otimes I$ and $P\otimes I$ can be written as a product of gates $K$, $K^{\dagger}$ and $W_v$; the same is easily seen to hold for $I\otimes H$ and $I\otimes P$. Consequently:
\ben[(i)]
\im  
As CZ $\propto  [P\otimes P] W_v$, the CZ gate can also be written as a product of $K$, $K^{\dagger}$ and $W_v$ gates; 
\im 
As $W^{\dagger}_h \propto HP^3H$, the gate $V = \bar{W}_h^{\dagger}K$ (regarded as a two--qubit operation acting nontrivially on either the first or second qubit) can be written as a product of $K$, $K^{\dagger}$ and $W_v$ gates as well;  
\im 
As the gate $K$ does not have finite order, there does not exist any integer $m$ such that $K^m= K^{\dagger}$. However, it is well known that for every $\delta>0$ there exists an integer $m=$ poly$(1/\delta)$ such that the distance (in operator norm) between $K^m$ and $K^{\dagger}$ is at most $\delta$.
\een
This shows that the gates $H$, $V$ as well as the CZ gate can be approximated with accuracy $\epsilon$ by a product of poly$(1/\epsilon)$ gates $K$ and $W_v$. Moreover, it is known that the CZ, $H$ and $V$ gates generate a dense subgroup of $U(4)$ (up to global phases). This hence proves Lemma \ref{lem_ising}.
\finpr

The proof of Lemma \ref{univ-Ising-gate} is now obtained as follows. We simultaneously conjugate the gates $K$ and $W_v$ with the operation $V^{\frac{1}{2}}$ acting on each qubit, yielding  
\be
\ba{rcl}
[V^{\frac{1}{2}}]K [V^{\frac{1}{2}}]^{\dagger} &=& T \\
\left[V^{\frac{1}{2}} \otimes V^{\frac{1}{2}} \right] W_v [V^{\frac{1}{2}} \otimes V^{\frac{1}{2}}]^{\dagger}\ &=&W_v\, .
\ea \label{conjugation} 
\ee  
As $K$ and $W_v$ generate a dense subgroup of $U(4)$ up to global phases due to Lemma \ref{lem_ising}, the same holds for $T$ and $W_v$ since the latter gates are obtained by conjugating the former with the same fixed local unitary operation. Finally, it follows that also the gates $T$ and $T^{\otimes 2} W_v T^{\otimes 2}$ generate a dense subgroup of $U(4)$ up to global phases: indeed, $T^{\dagger}$ can be approximated with arbitrary accuracy by powers of $T$, such that $W_v = T^{\dagger}[T^{\otimes 2} W_v T^{\otimes 2}] T^{\dagger}$ can be approximated with arbitrary accuracy by products of $T$ and  $T^{\otimes 2} W_v T^{\otimes 2}$.

We have thus showed that every two--qubit operation can be approximated with arbitrary accuracy by products of $T$ and $T^{\otimes 2} W_v T^{\otimes 2}$. The Solovay--Kitaev Theorem then guarantees fast convergence, i.e.~every poly--size $n$--qubit circuit composed of two--qubit gates can be approximated with accuracy $\epsilon$ by a circuit of poly$(1/\epsilon, n)$  gates $T$ and $T^{\otimes 2} W_v T^{\otimes 2}$; moreover, the latter circuit can be found efficiently, i.e.~in  poly$(1/\epsilon, n)$ time. Hence this proves the claim.

\section{Hadamard rotation with Potts--type gates}
\lb{app:HadamardPotts}

In the following we show how to obtain a Hadamard gate with Potts--type gates. 
Consider a general state (defined at the logical level), $|\bs{\psi}\ra =\alpha |\bs{0}\ra + \beta |\bs{1}\ra $, where $\alpha$ and $\beta$ are arbitrary, normalized coefficients, $|\alpha|^2 +|\beta|^2=1$. Our goal is to apply a Hadamard gate to this state, that is, to transform it to $|\bs{\psi}\ra :=\alpha |\bs{+}\ra + \beta |\bs{-}\ra $, where
\be
|\bs{\pm}\ra := \frac{1}{\sqrt{2}} (|\bs{0}\ra \pm |\bs{1}\ra)  \, ,
\lb{eq:logical+-}
\ee
by analogy with the conventional definitions $|\pm\ra := (|0\ra \pm|1\ra)/\sqrt{2}$. 

To apply this gate, we first observe that a Hadamard rotation can be decomposed as
\be
\bs{H} = \bs{R_z}(3\pi/2) \bs{R_x}(\pi/4) \bs{R_z}(3\pi/2)\, .
\ee
Then these three gates can be achieved with the following sequence of Potts--type gates (see Fig.~\ref{fig:hadamard-Potts}):
\ben
\im
A two qutrit--gate between an auxiliary qutrit in the state $|2\ra$ and the lower qutrit of the logical qubit with $(\mu,\nu)=( -i,1)$,
which results in the state $\alpha |0\ra|1\ra - iÊ\beta |1\ra|2\ra$. 
\item  
A single--qutrit gate on the upper qutrit with $(\mu,\nu)= (-i,1)$, and then a single--qutrit gate on the second qutrit with $(\mu,\nu)= (1,1)$, which results in the state  
$\alpha (-i|0\ra + |1\ra + |2\ra)(|0\ra+|1\ra + |2\ra) -i\beta (|0\ra -i |1\ra + |2\ra)(|0\ra+|1\ra + |2\ra)$. 
\item 
A two--qutrit gate between an auxiliary qutrit in $|2\ra$ and the upper qutrit with $(\mu,\nu)=(0,1)$, and the same gate between an auxiliary qutrit in $|1\ra$ and the lower qubit, which results in the state 
$\alpha (-i|0\ra + |1\ra)(|1\ra + |2\ra) -i\beta (|0\ra - i |1\ra)(|1\ra +|2\ra)$.
\item 
A two qutrit gate between the upper and lower qutrit with $(\mu,\nu)=(0,1)$, which yields the state 
$\alpha (-i|0\ra |1\ra -i|0\ra|2\ra + |1\ra|2\ra) -i\beta (|0\ra|1\ra + |0\ra|2\ra -i|1\ra|2\ra)$.
\item 
\lb{im:epsilon}
A two qutrit gate between an auxiliary qutrit in $|0\ra$ and the upper qutrit with $(\mu,\nu)=(\epsilon,1)$, and the same gate between an auxiliary qutrit in $|2\ra$ and the lower qutrit. 
Then a two--qutrit gate between an auxiliary qutrit in $|1\ra$ and the upper qutrit with $(\mu,\nu)=(1/(\epsilon\sqrt{2}),1)$, and the same gate between an auxiliary qutrit in $|1\ra$ and the lower qutrit. 
Up to terms of order $\mc{O}(\epsilon^2)$, the resulting state is
  $[\alpha(-i|0\ra|1\ra + |1\ra|2\ra) -i \beta(|0\ra |1\ra -i |1\ra|2\ra)]/\sqrt{2}$.
 See below for a discussion of the accuracy of the gate due to neglecting these higher order terms. 
 \item 
A two--qutrit gate between an auxiliary qutrit in $|0\ra$ and the upper qutrit with $(i,1)$, and another two--qutrit gate between an auxiliary qutrit in $|1\ra$ and the upper qutrit. The resulting state is
$[\alpha (|0\ra|1\ra + i |1\ra|2\ra) + \beta (|0\ra|1\ra - i |1\ra|2\ra)]/\sqrt{2}$.

\item 
Finally, a two--qutrit gate between an auxiliary qutrit in $|2\ra$ and the lower qutrit with $(-i,1)$, after which the state of the system is
$[\alpha (|0\ra|1\ra + |1\ra|2\ra) + \beta (|0\ra|1\ra -  |1\ra|2\ra)]/\sqrt{2} 
=\alpha (|\bs{0}\ra +|\bs{1}\ra )/\sqrt{2} + \beta  (|\bs{0}\ra -|\bs{1}\ra)/\sqrt{2} 
= \bs{H} |\bs{\psi}\ra 
$,
as we wanted to show.
\een

Note that, due to step \ref{im:epsilon}, the desired gate is achieved up to an error of $\mc{O}(\epsilon^2)$, where we use e.g.~the Jamio\l kowski Fidelity $F$ \cite{Gi05} as a measure for the gate fidelity. Making use of the distance measure between the imperfect operation and the ideal Hadamard gate, $D({\cal E}, {\cal U})=\sqrt{1-F}$, together with the chaining inequality \cite{Gi05,Du08}, $D({\cal E}_1\circ{\cal E}_2,{\cal U}_1\circ{\cal U}_2) \leq D({\cal E}_1, {\cal U}_1) + D({\cal E}_2, {\cal U}_2)$, we conclude that the total distance after $M$ applications of an imperfect operation of this kind in a quantum circuit is at most $M$ times the distance of a single imperfect operation. That is, $\sqrt{1-F_{\rm tot}} \leq M \sqrt{1-F}$, where $F_{\rm tot}$ denotes the Jamio\l kowski fidelity of the overall circuit containing $M$ imperfect operations, each with Jamio\l kowski fidelity $F$. Hence, for any circuit consisting of a polynomial number of gates, $M=\mc{O}(N)$, a polynomial accuracy in $\epsilon$, that is $\epsilon=\mc{O} (1/{\rm poly} (N))$, suffices to achieve a polynomial accuracy of the total circuit. Notice that this is sufficient to obtain a final approximation of the partition function with polynomial accuracy
\footnote{
One may also use results from fault--tolerant quantum computation, which state that also noisy gates with a small but constant error suffice to perform quantum computation. This implies that $\epsilon$ can be a small constant not depending on the size of the circuit, however additional overhead for encoding and fault-tolerant gate implementation would be required in this case.}.

\bfi[htb]
\centering
\psfrag{i}{\small{$|\bs{\psi}\ra$}}
\psfrag{2}{\small{$|2\ra$}}
\psfrag{0}{\small{$|0\ra$}}
\psfrag{1}{\small{$|1\ra$}}
\psfrag{a}{\small{$(-i,1)$}}
\psfrag{c}{\small{$(-i,1)$}}
\psfrag{e}{\small{$(1,1)$}}
\psfrag{A}{\small{$(0,1)$}}
\psfrag{C}{\small{$(0,1)$}}
\psfrag{E}{\small{$(0,1)$}}
\psfrag{H}{\small{$(\epsilon,1)$}}
\psfrag{J}{\small{$(\epsilon,1)$}}
\psfrag{F}{\small{$1$}}
\psfrag{L}{\small{$(\frac{1}{\epsilon\sqrt{2}},1)$}}
\psfrag{N}{\small{($\frac{1}{\epsilon\sqrt{2}},1)$}}
\psfrag{P}{\small{$(i,1)$}}
\psfrag{R}{\small{$(i,1)$}}
\psfrag{T}{\small{$(-i,1)$}}
\psfrag{o}{\small{$\bs{H}|\bs{\psi}\ra$}}
\psfrag{x}{\small{$\bs{R_z}(3\pi/2)$}}
\psfrag{y}{\small{$\bs{R_x}(\pi/4)$}}
\psfrag{z}{\small{$\bs{R_z}(3\pi/2)$}}
\psfrag{t}{\small{time}}
\includegraphics[width=1.05\columnwidth]{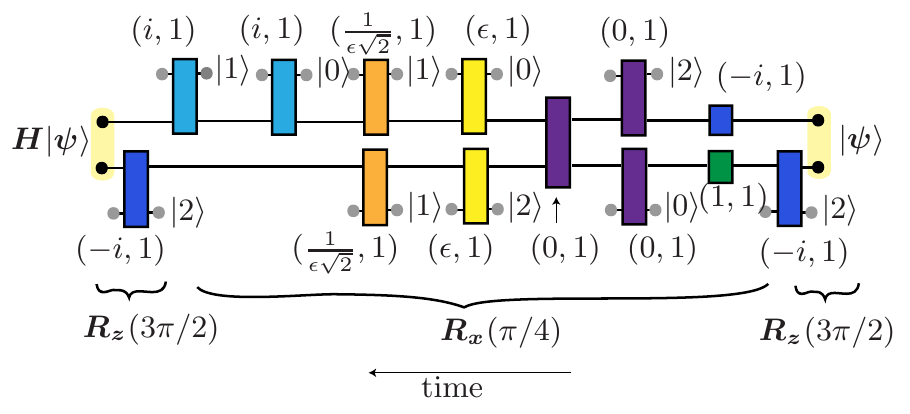}
\caption{
The Hadamard rotation at the logical level, $\bs{H}$, is achieved by first performing a $z$--rotation $\bs{R_z}$, then an $x$--rotation $\bs{R_x}$ (which is itself composed of various gates) and finally another $z$--rotation $\bs{R_z}$. 
Most of the gates involve auxiliary qubits, whose input state is indicated on the right (since time runs from right to left).
See the caption of Fig.~\ref{fig:I1andP} for the meaning of the pair of numbers and the color associated to each gate. See the text for the transformation of the state $|\bs{\psi}\ra$ at each stage of the circuit. 
}
\label{fig:hadamard-Potts}
\efi

\section{Hadamard rotation with $\mathbb{Z}_2$ Lattice Gauge Theory--type gates}
\lb{app:HadamardLGT}

Here we show how to general single--qubit rotation of \eqref{eq:Ualphabetagamma} with $\mathbb{Z}_2$ LGT--type gates. 
This operation includes $\bs{R_z}$ rotations, which we have shown how to perform in Sec.~\ref{ssec:3DZ2LGT}, and the non--normalized version of the logical Hadamard gate $\tilde{\bs{H}}$. 
We will implement the latter as a teleportation--based gate. 
Intuitively, the idea is the following. 
To apply $\tilde{\bs{H}}$ to the logical qubit $|\bs{\psi}\ra$, we prepare two logical qubits next to it in the state $|\bs{0}\bs{+}\ra + |\bs{1}\bs{-}\ra$ (see \eqref{eq:logical+-}). 
Then we apply a projection onto a Bell state $|\bs{0}\bs{0}\ra + |\bs{1}\bs{1}\ra$ between the qubit we want to teleport and the first of these two logical qubits.
The third logical qubit is then left in the desired state, namely $\tilde{\bs{H}}|\bs{\psi}\ra$. 

More precisely, consider a general initial logical state $|\bs{\psi}\ra =a |\bs{0}\ra + b |\bs{1}\ra$, where $a$ and $b$ are unknown, normalized coefficients $|a|^2+|b|^2=1$.
Then we transform this state into $\bs{R_z}(\pi/2)\tilde{\bs{H}}\bs{R_z}(\alpha)|\bs{\psi}\ra$ by applying the following sequence of steps (see Fig.~\ref{fig:Hadamard-LGT}):
\ben
 \im 
Single--qubit gates with $e^{\beta J} =1$ to all physical qubits of the two logical qubits on the right.
Each of these gates transforms the state of the physical qubits from $|0\ra$ to $|0\ra + |1\ra =: \sqrt{2} |+\ra$. Thus, this results in the state $|\bs{\psi}\ra 16 |++++\ra^{\otimes 2} $. 
\im
The four--qubit gate with $e^{\beta J}=e^{i(\alpha +\pi/2)}$ on the top physical qubit of the first logical qubit. 
Also the four--qubit gate with $e^{\beta J}=0$ inside the plaquette where the last two logical qubits are defined, and the four--qubit gates with $e^{\beta J}=0$ on the physical qubits on top and on the right of the last two logical qubits (as indicated in the second step of Fig.~\ref{fig:Hadamard-LGT}).
This results in the state 
$(a|\bs{0}\ra +e^{i(\alpha +\pi)} b|\bs{1}\ra )
[|0000\ra+|0011\ra+|1100\ra+|1111\ra]^{\otimes 2}$. 
\im
The four--qubit gates with $e^{\beta J}=0$ between the physical qubits on the right and on the bottom of the last two logical qubits (as indicated in the third step of Fig.~\ref{fig:Hadamard-LGT}).
The resulting state is 
$(a|\bs{0}\ra +e^{i(\alpha +\pi)} b|\bs{1}\ra ) 
[|\bs{0}\ra + |\bs{1}\ra]^{\otimes 2}$
\im
The four--qubit gate with $e^{\beta J}=i$ applied on the top physical qubit of the second logical qubit. 
A four--qubit gate with $e^{\beta J}=0$ between the physical qubit on the right of the second logical qubit and the physical on the right of it, which we refer to as the ``middle qubit'' (see fourth step of Fig.~\ref{fig:Hadamard-LGT}), which propagates the state of the first qubit to the middle one. 
Then the four--qubit gate with $e^{\beta J}=i$ between the middle qubit and the left physical qubit of the third logical qubit. 
The resulting state is
$(a|\bs{0}\ra +e^{i(\alpha +\pi)} b|\bs{1}\ra ) 
[|\bs{0}\ra(|\bs{0}\ra + i|\bs{1}\ra)  -|\bs{1}\ra (|\bs{0}\ra  - i|\bs{1}\ra]$.
\im 
Finally, two concatenated four--qubit gates with $e^{\beta J}=0$ between the first and the second logical qubit, which corresponds to the projection onto the Bell state $\la \bs{0}|\la \bs{0}| + \la \bs{1}|\la \bs{1}|$.
We are interested in the state of the third logical qubit, since this contains the result of the teleportation. 
This state is $a (|\bs{0}\ra + i|\bs{1}\ra) +e^{i\alpha} b( |\bs{0}\ra - i |1\ra)$, which equals 
\be
\bs{R_z}(\pi/2)\tilde{\bs{H}}\bs{R_z}(\alpha) |\bs{\psi}\ra\, .
\lb{eq:Rzpi2}
\ee
\een

Thus, this sequence of gates applies essentially the first two logical gates of a general single qubit rotation of \eqref{eq:Ualphabetagamma}.
The rotation $\bs{R_z}(\pi/2)$ appearing in \eqref{eq:Rzpi2} can be compensated in the next rotation (marked as time step 8 in Fig.~\ref{fig:normalform}) by applying a four--qubit gate with $e^{i(\beta + \pi/2)}$ (instead of $e^{i(\beta + \pi)}$, which would be the analogous case to the rotation of step 3, where $e^{i (\alpha + \pi)}$ was applied). This results in the state $\bs{R_z}(\beta)\tilde{\bs{H}}\bs{R_z}(\alpha) |\bs{\psi}\ra$. 
Then a teleportation--based Hadamard gate $\tilde{\bs{H}}$ is applied with the same procedure as the one described above,
and this gives rise to the state $\bs{R_z}(\pi/2)\tilde{\bs{H}}\bs{R_z}(\beta)\tilde{\bs{H}}\bs{R_z}(\alpha)$. 
The final rotation must correct for this phase, and the four--qubit gate at step 11 of Fig.~\ref{fig:normalform} must have $e^{i(\gamma-\pi/2)}$ (instead of $e^{i\gamma}$). The overall sequence implements the desired general single qubit gate, $\bs{R_z}(\gamma) \tilde{\bs{H}}\bs{R_z}(\beta) \tilde{\bs{H}} \bs{R_z}(\alpha)|\bs{\psi}\ra$.

\bfid[htb]
\centering
\psfrag{b}{\small{$e^{\beta J} \! =\! i$}}
\psfrag{c}{\small{$e^{\beta J}\! =\! e^{i(\alpha + \pi)}$}}
\psfrag{p}{\small{$|\bs{\psi}\ra$}}
\psfrag{1}{\small{$|\bs{0}\ra$}}
\psfrag{2}{\small{$|\bs{0}\ra$}}
\psfrag{a}{\small{$e^{\beta J} \! =\! 1$}}
\psfrag{d}{\small{$e^{\beta J}\! =\! 1/\sqrt{2}$}}
\psfrag{e}{\small{$e^{\beta J} \! =\! 0$}}
\psfrag{Z}{\small$\bs{R_z}(\pi/2)\tilde{\bs{H}}\bs{R_z}(\alpha) |\bs{\psi}\ra$}
\psfrag{t}{time}
\psfrag{P}{processing of qubits}
\psfrag{x}{\small{$x$}}
\psfrag{y}{\small{$y$}}
\psfrag{z}{\small{$z$}}
\includegraphics[width=1.85\columnwidth]{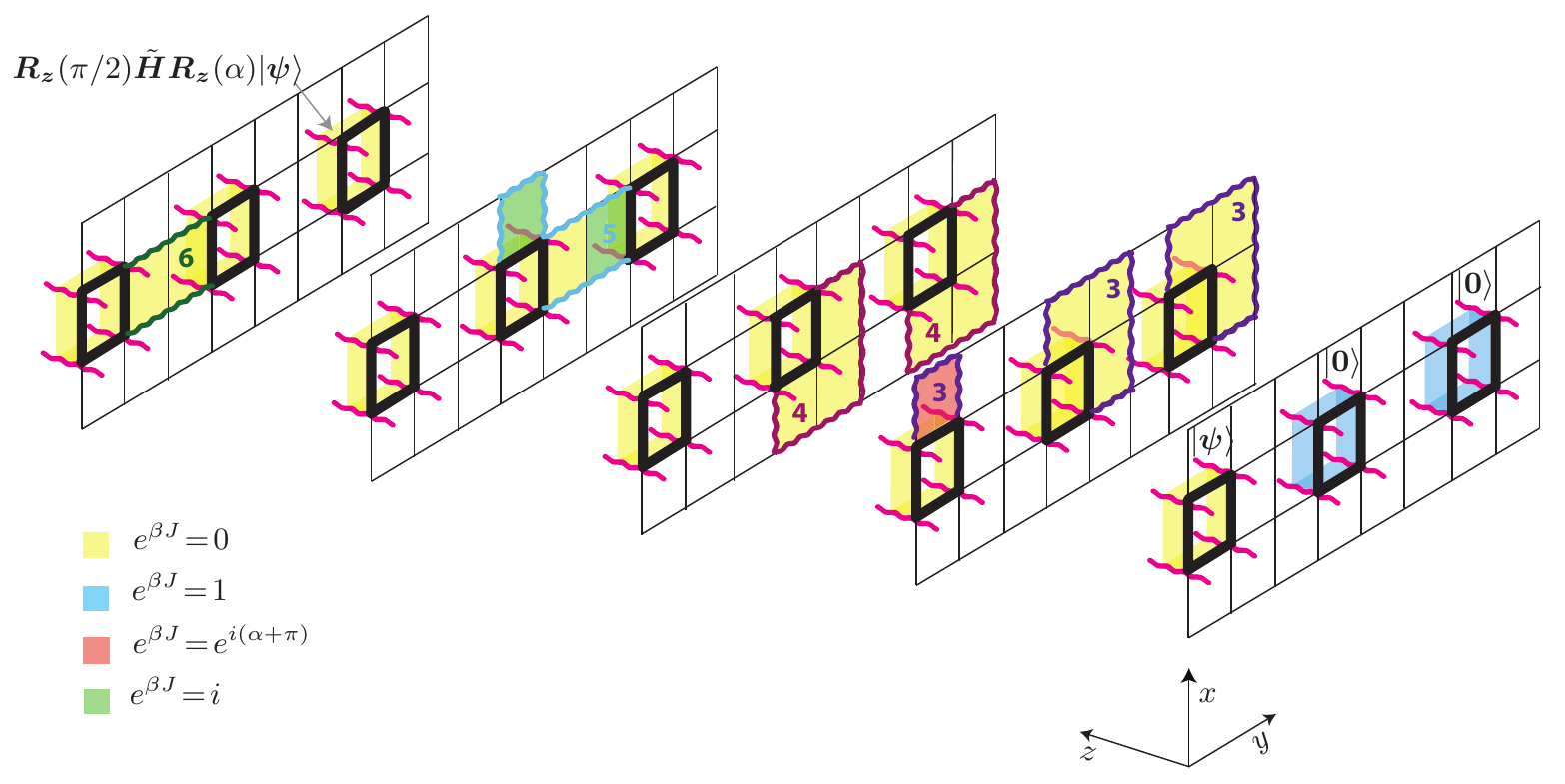}
\caption{Part of a circuit with $\mathbb{Z}_2$--LGT--type gates that shows the processing of a logical qubit  $|\bs{\psi}\ra$. First, a logical rotation $\bs{R_z}(\alpha)$ is applied to it. Then, a logical teleportation--based Hadamard gate (up to a phase) is applied, which results in the state $\bs{R_z}(\pi/2)\tilde{\bs{H}}\bs{R_z}(\alpha) |\bs{\psi}\ra$ (see text).
Note that the qubit is processed in the $z$ direction (corresponding to time) as well as the $y$ direction.
Colored faces indicate faces where the coupling strength has a specific value, as indicated in the legend. Wavy lines indicate edges whose spins have been fixed by the gauge, and their color has no meaning nor a correlation with the color of the faces. They are a guide to the eye to verify that no loops are formed, as shown in Fig.~\ref{fig:normalform}.
}
\lb{fig:Hadamard-LGT}
\efid

Moreover, we apply single--qubit identities at the physical level $I_1:=|0\ra\la 0| + |1\ra\la 1|$ on all physical qubits which belong to a logical qubit unless otherwise specified, as mentioned in Sec.~\ref{ssec:3DZ2LGT} (yellow faces in the temporal direction in Fig.~\ref{fig:Hadamard-LGT}). This gate is achieved by first rescaling the face interaction energy (Eq.~\eqref{eq:int-LGT}) to $-(J_f-1)\delta(s_i+ s_j + s_k +s_l)$. According to \eqref{eq:singlequbitgate} this gives rise to single qubit gate whose diagonal elements are 1 and the off--diagonal elements are $e^{\beta (J_f-1)}$. Then the gate is achieved by setting $e^{\beta (J_f-1)}= \zeta$, where $\zeta$ is a polynomially small number. By the same argument as for the Potts single--qubit identity (see Sec.~\ref{ssec:Potts} remark \eqref{im:Potts-remark-infinity}), this will result in polynomial error at the end of the computation, which is within the accuracy of the result.

Note that the unnormalized gate $\tilde{\bs{H}}$ is applied once in every volume of 4 units in the $x$ direction, 12 units in the $y$ direction, and 7 units in the $z$ direction. As mentioned in Sec.~\ref{ssec:3DZ2LGT}, the number of  these blocks is what gives rise to the factor appearing in \eqref{eq:Z-LGT}. 

Since the general single qubit unitary gate involves a number of spins fixed by the gauge, it is important to verify that no closed loops of fixed spins are formed~\cite{Cr77}. 
This is because the gauge fixing only yields an equivalent theory if no loops are formed, since, e.g., order parameters of the theory are defined as products of variables on such closed loops. 
To verify that no such loops are formed we project our 3D lattice into a 2D lattice along the time direction (see Fig.~\ref{fig:normalform}), where one can easily recognize that there is no closed path of edges fixed by the gauge. 

\bfid[htb]
\centering
\psfrag{a}{\small{$e^{i(\alpha + \pi)}$}}
\psfrag{H}{\small{$\tilde{\bs{H}}$}}
\psfrag{b}{\small{$ e^{i(\beta + \pi/2)}$}}
\psfrag{c}{\small{$e^{i(\gamma - \pi/2)}$}}
\psfrag{N}{\small{Two--qubit gate}}
\psfrag{S}{\small{General single--qubit gate $\bs{R_{z}}(\gamma) \tilde{\bs{H}} \bs{R_z}(\beta) \tilde{\bs{H}} \bs{R_z}(\alpha)$}}
\psfrag{s}{\small{all time steps}}
\psfrag{t}{\small{$\bs{e^{i\varphi \sigma_z\otimes \sigma_z}}$}}
\psfrag{r}{\small{processing of qubits}}
\psfrag{T}{	\small{time}}
\psfrag{x}{\small{$x$}}
\psfrag{y}{\small{$y$}}
\psfrag{z}{\small{$z$}}
\includegraphics[width=1.95\columnwidth]{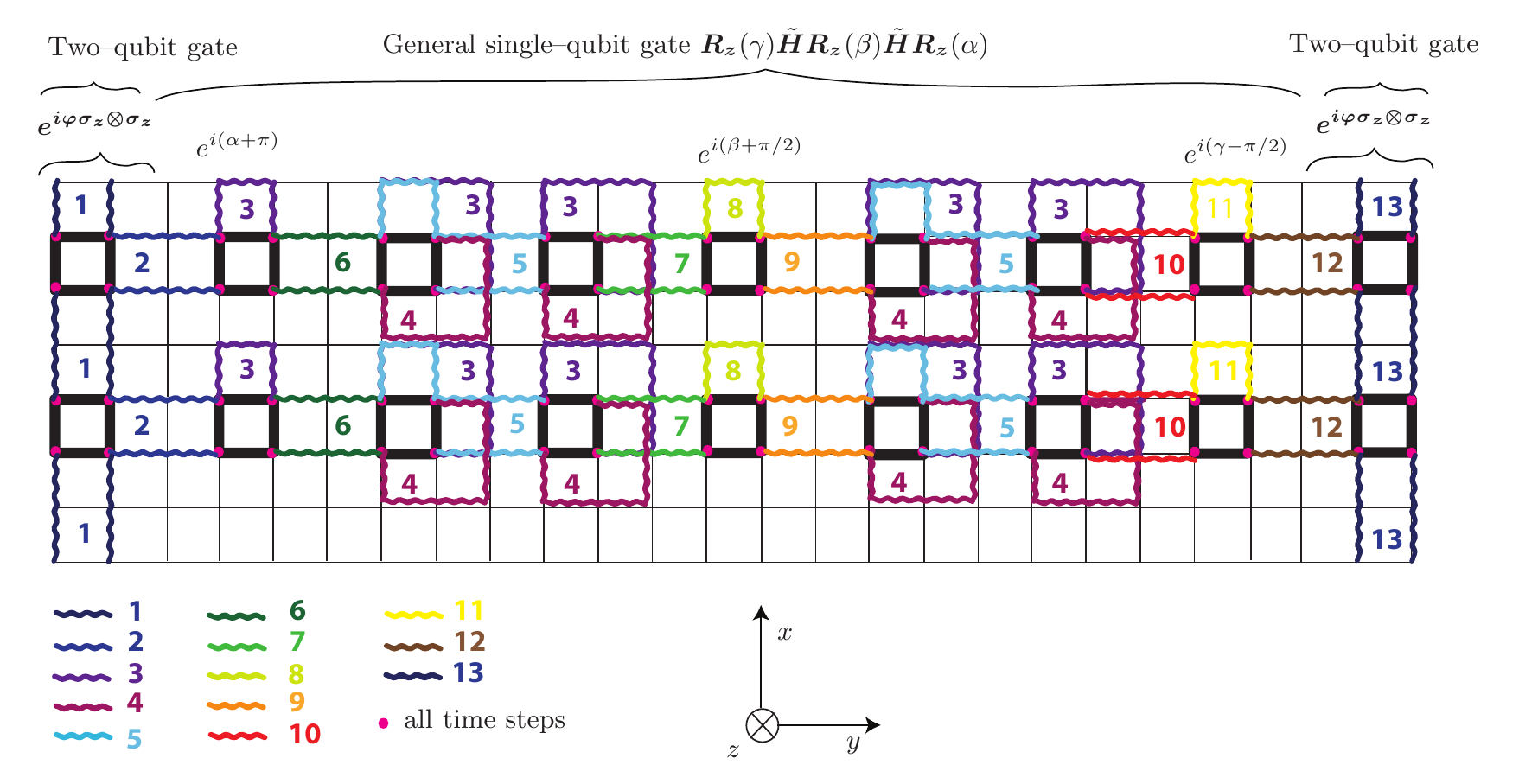}
\caption{
A projection of the circuit of the 3D $\mathbb{Z}_2$ LGT into its spatial dimension. 
Part of the circuit in its normal form (that is, in the structure two--qubit gate, single--qubit gate, two--qubit gate) is illustrated.
Logical qubits are indicated with thick, black lines. 
All edges in the time direction (going out of the paper) which are boundary to a logical qubit are fixed by the gauge (pink dots). Edges in the spatial direction are either not fixed by the gauge (black, thin lines), or are fixed by the gauge at \emph{different} time steps $1 \ldots 13$ indicated with a wavy, colored edges. Hence, a loop of edges fixed by the gauge would correspond in this figure to a closed loop of wavy, colored edges. It can be verified by inspection that no such loops are present.
}
\lb{fig:normalform}
\efid

\bibliographystyle{apsrev}
\bibliography{/Users/GemmaDelasCuevas/Documents/Professional/PhD/Special-files/all-my-bibliography}

\end{document}